\documentclass[10pt]{rspublic}
\usepackage{graphicx}

\newtheorem{thm}{Theorem}
\newtheorem{lem}{Lemma}
\theoremstyle{definition}

\theoremstyle{definition}

\theoremstyle{remark}

\theoremstyle{plain}

\newcommand{\Q}{{\mathbb{Q}}}

\newcommand{\Z}{{\mathbb{Z}}}

\newcommand{\myfrac}[2]{{#1}/{#2}}

\title{Uniform asymptotics of the coefficients of unitary moment polynomials}
\author{G.A. Hiary and M.O. Rubinstein
\footnote{Both authors are supported by the
National Science Foundation under awards DMS-0757627 (FRG grant) and
DMS-0635607. In addition, the second author is supported by an NSERC Discovery Grant.}
}

\begin{document}
\maketitle

\begin{abstract}{uniform asymptotics, unitary moment polynomials}
Keating and Snaith showed that the $2k^{th}$ absolute moment of the characteristic
polynomial of a random
unitary matrix evaluated on the unit circle is given by a polynomial of degree $k^2$. In this article,
uniform asymptotics for the coefficients of that  polynomial are derived, and a
maximal coefficient is located. Some of the asymptotics are given in explicit
form. Numerical data to support these calculations are presented. Some
apparent connections between random matrix theory and the Riemann zeta function
are discussed.
\end{abstract}

\section{Introduction}

Let $Z_N(A,\theta)$ denote the characteristic polynomial of an $N\times N$
unitary matrix $A$ evaluated at $\exp(i\theta)$.
Let $\theta_n$ denote the eigenphases of $A$. Then,
    $Z_N(A,\theta):=\mathrm{det}\, (e^{i\theta} I-A)= \prod_n (e^{i\theta}-e^{i\theta_n})$,
Keating and Snaith~\cite{KS} gave an explicit expression for the expected moments of
$|Z_N(A,\theta)|$:
\begin{eqnarray}
    \label{eq:KS moments}
    \mathbb{E}_N |Z_N(A,\theta)|^{2k}=\prod_{j=1}^N
    \frac{\Gamma(j)\Gamma(j+2k)}{\Gamma(j+k)^2}\,,
    \quad \Re{k>-1/2}.
\end{eqnarray}
The expectation is taken with respect to the normalized Haar measure on the
group of $N\times N$ unitary matrices $U(N)$. This can also be written
in terms of the Barnes $G$-function, which is the entire function
of order 2 defined by
    $G(z+1)= (2\pi)^{z/2} e^{-(z+(1+\gamma) z^2)/2} \prod_{n=1}^\infty (1+z/n)^n e^{-z+z^2/(2n)}$,
and satisfying the relations $G(1) = 1$, $G(z+1)=\Gamma(z)G(z)$.
The r.h.s. of \eqref{eq:KS moments} equals
\begin{eqnarray} \label{eq:temp1}
    \frac{G(k+1)^2}{G(2k+1)}\,\frac{G(N+1)G(N+2k+1)}{G(N+k+1)^2}\,.
\end{eqnarray}
When $2k \in \Z$, this simplifies
to a polynomial in $N$ of degree $k^2$. This polynomial can be expressed
in several equivalent ways, each of which is useful:
\begin{eqnarray}
    \label{eq:pkxeq}
    &&\mathbb{E}_N |Z_N(A,\theta)|^{2k}
    = \frac{G(k+1)^2}{G(2k+1)}\,
    \prod_{0 \leq i,j \leq k-1} (N+i+j+1) \\
    &=& \prod_{j=0}^{k-1} \frac{j!}{(j+k)!}
    \prod_{j=1}^k (N+j)^j \prod_{j=k+1}^{2k} (N+j)^{2k-j}
    =
    \prod_{0 \leq i,j \leq k-1} \left(1+\frac{N}{i+j+1} \right). \notag
\end{eqnarray}
Expanding, we define,
\begin{eqnarray}
    \mathbb{E}_N |Z_N(A,\theta)|^{2k}=:\sum_{r=0}^{k^2} c_{r}(k) N^{k^2-r}=:P_k(N)\,.
\end{eqnarray}
The asymptotics of the leading coefficient $c_{0}(k)$
can be obtained from the asymptotics of the Barnes function:
    $\log G(z+1) \sim
    z^2 (\log(z)/2 -3/4) + \log(2\pi)z/2 -\log(z)/12 +\zeta'(-1) +O(1/z)$,
from which it follows that
$c_0(k) = -k^2\log{k} -k^2\log{4} +3k^2/2 -\log(k)/12+\zeta'(-1) +O(1/k)$.

\subsection{Results}

The main purpose of this paper is to develop {\em{uniform}} asymptotics in $r$
for the coefficients $c_{r}(k)$, $0\le r\le k^2$, as $k\to \infty$. We also
obtain some explicit asymptotics, and estimate in $r$, for given $k$,
the maximal $c_r(k)$.

We prove the following theorem in the next two sections as a sequence of five lemmas:
\begin{thm}
    \label{thm:1}
    Let $u:=u_{r,k}$ be the unique positive number satisfying
        $u\sum_{j=1}^k \frac{j}{u+j} + u\sum_{j=k+1}^{2k} \frac{2k-j}{u+j} = k^2-r$.
    Further, define $U:=U_{r,k}$ by
        $U=u \sum_{j=1}^k \frac{j^2}{(u+j)^2}+u\sum_{j=k+1}^{2k} \frac{j(2k-j)}{(u+j)^2}$.
    Then, uniformly in $r\in (0,k^2)$, and as $k\to \infty$,
    \begin{eqnarray}
        \label{eq:uniform}
        c_r(k)=\left(\frac{r}{k^2} \right)^r
        \binom{k^2}{r}
        \frac{P_k(u)}{\sqrt{U}\, u^{k^2-r}}
        \left(1-\frac{r}{k^2}\right)^{k^2-r+1/2}\,r^{1/2}\left[1+O\left(\frac{\log k}{k^{2/3}}\right) \right]\,.
    \end{eqnarray}
    Furthermore, if we fix $\alpha<1$, then, uniformly
    in $0 \leq r \leq k^{\alpha}$,
    \begin{eqnarray}
        \frac{c_{r}(k)}{c_0(k)}=\frac{k^{3r}}{r!} \left[1+O(k^{2(\alpha-1)})
        \right]= \binom{k^2}{r} k^r \left[1+O(k^{2(\alpha-1)}) \right]\,.
    \end{eqnarray}
    and
    \begin{eqnarray}
        c_{k^2-r}(k) =\frac{(kA)^{r}}{r!} \left[1+ O(k^{2(\alpha-1)}\log{k})\right]=
        \binom{k^2}{r}\frac{A^r}{k^r}\left[1+O(k^{2(\alpha-1)}\log{k}) \right]
    \end{eqnarray}
    with the implied constants in the $O$ terms depending on $\alpha$ and where
    \begin{eqnarray}\label{eq:adef}
        A=2\sum_{j=k+1}^{2k} 1/j=\log 4+\frac{1}{2k}+O(1/k^2)
    \end{eqnarray}
\end{thm}

Regarding the location of the maximal $c_r(k)$ we prove in section~\ref{sec:max coeff}:
\begin{thm}
\label{thm:2}
Let
\begin{eqnarray}
\mu:= \sum_{j=1}^k \frac{j}{j+1}+\sum_{j=k+1}^{2k} \frac{2k-j}{j+1}= k\log 4-\log (k/2)+1/2-\gamma +O(1/k)\,,
\end{eqnarray}
where $\gamma=0.57721\ldots$ is the Euler constant. Then, there exists $\rho>0$ such that,
for all $k$ sufficiently large, a maximal $c_r(k)$ occurs for some
\begin{eqnarray} \label{eq:maxint}
r\in [k^2-\mu-\rho \log(k)^2/k ,k^2-\mu+1+\rho \log(k)^2/k]\,,
\end{eqnarray}
and no maximal $c_r(k)$ occurs outside of that interval.
\end{thm}
Notice that the size of the interval in~\eqref{eq:maxint} is slightly larger
than 1. So typically, it will only contain one integer, and the maximum
with occur for $r=\lceil k^2-\mu \rceil$,
but it can
contain two integers in the event that $\mu$ is very close to an integer.

In section~\ref{sec:more precise} we derive a more precise formula for
the leading coefficients, valid for $r=O(k^\alpha)$ with $\alpha <2$,
the first few terms of which are listed below:

\begin{thm}
\label{thm:3}

\begin{eqnarray}
    \label{eq:thm3}
    \frac{c_r(k)}{c_0(k)} &=& \frac{k^{3r}}{r!} \exp \Biggl(
        -\frac {7r(r-1)}{12k^2}
        -\frac {r(r-1)(26r-15)}{144k^4}
        -\frac {r(r-1)(583r^2-715r+183)}{6480k^6} \notag \\
        &-&\frac {r(r-1)(2758r^3-4499r^2+463r+1491)}{51840k^8} -\ldots
        - \frac{q_j(r)}{k^{2j}}-\ldots
    \Biggr)\,.
\end{eqnarray}
The $j$-th term in the exponent is of the form $q_j(r)/k^{2j}$, where
$q_j(r)$ is a polynomial in $\Q[r]$ of degree $j+1$, divisible by $r(r-1)$,
and satisfying the bound $q_j(r)=O(j \log(j) r(\lambda r)^j)$, for
some $\lambda>0$.

\end{thm}

The above expansion can be transformed into a similar one involving
the binomial coefficient $\binom{k^2}{r}$:
\begin{eqnarray}
    \label{eq:cr binom}
    \frac{c_r(k)}{c_0(k)} &=& k^r \binom{k^2}{r} \exp \Biggl(
        -\frac {r(r-1)}{12k^2}
        -\frac {r(r-1)(2r-3)}{144k^4}
        -\frac {r(r-1)(43r^2-175r+183)}{6480k^6} \notag \\
        &-&\frac {r(r-1)(166r^3-611r^2+31r+1059)}{51840k^8} -\ldots
        - \frac{\tilde{q}_j(r)}{k^{2j}} -\ldots
    \Biggr)\,.
\end{eqnarray}
This can be achieved by writing
    $\binom{k^2}{r}\frac{r!}{k^{2r}}
    = \prod_{j=1}^{r-1} \left(1-\frac{j}{k^2}\right)
    = \exp (-\sum_{j=1}^{r-1}
      \sum_{m=1}^\infty \frac{j^m}{mk^{2m}}
    )
    =
    \exp(
      - \sum_{m=1}^\infty \frac{B_{m+1}(r)-B_{m+1}(1)}{(m+1)mk^{2m}}
    )$,
where $B_m(x)$ are the Bernoulli polynomials, and we have used~\eqref{eq:bernoulli}.

The apparent smaller coefficients in the exponent of~\eqref{eq:cr binom}
in comparison to~\eqref{eq:thm3}
suggests that the binomial coefficient appears naturally as a factor in the asymptotics. A heuristic
explanation is provided in the next section.

It turns out the problem of deriving uniform asymptotics for the $c_r(k)$'s is
similar to that of deriving uniform asymptotics for the unsigned Stirling
numbers of the first kind. These numbers are usually denoted by $S(k,r)$, and
can be defined by the generating function
$\prod_{j=0}^{k-1} (x+j)=:\sum_{r=1}^k S(k,r) x^r$.
The above generating function bears some resemblance to the generating function
of the $c_r(k)$'s given in (\ref{eq:pkxeq}), which is the reason the settings
in the two problems are similar.

Our proof of lemma~\ref{lem:i2} parallels the
proof of asymptotic (1.3) in Moser and Wyman~\cite{MW}, where uniform
asymptotics for the $S(k,r)$'s were developed. See also Theorem 2 in the later
work of~\cite{CRT}, which relies on the work of~\cite{MW}.

\subsection{Motivation and connections to the Riemann zeta function}

For a fixed $k$ with $\Re{k} > -1/2$, and as $T \to \infty$, 
the
$2k^{th}$ moment of the Riemann zeta function on the critical line
$\zeta(1/2+it)$ is conjectured to have an asymptotic expansion:
$\int_0^T |\zeta(1/2+it)|^{2k}\, dt \sim \int_0^T \tilde{P}_k(\log(t/(2\pi)))dt$,
where $\tilde{P}_k(x)$ is an asymptotic series in descending powers of $x$,
of the form:
$\tilde{P}_k(x):=\sum_{r=0}^{\infty} \tilde{c}_r(k) x^{k^2-r}$.
When $x\in \Z$, then this series terminates and is a polynomial of
degree $k^2$ in $x$:
$\tilde{P}_k(x):=\sum_{r=0}^{k^2} \tilde{c}_r(k) x^{k^2-r}$.

The leading coefficient is expected to be of the form $ \tilde{c}_0(k)= a_k
g_k/k^2!$, where $a_k$ is a certain, generally understood, ``arithmetic
factor,'' and $g_k$ is an integer for integer values of $k$ (see~\cite{CG1}).
It is well-known $g_1=1$, $g_2=2$, and it was conjectured $g_3=42$,
$g_4=24024$, in~\cite{CG2} and~\cite{CGo}, respectively. But little was known
about $g_k$ in general.

Based on the analogous computation in random matrix theory,
Keating and Snaith~\cite{KS} conjectured that
$\myfrac{g_k}{(k^2!)}=\myfrac{G^2(k+1)}{G(2k+1)}$, or $c_0(k)=\tilde{c}_0(k)/a_k$.
In other words, the leading coefficient of $P_k(x)$ and of $\tilde{P}_k(x)$ are expected
to coincide, except for ``arithmetic effects.''

Recently, Conrey, Farmer, Keating, Rubinstein, and Snaith~\cite{CFKRS1}
conjectured, for integer $k$, a formula for the polynomial $\tilde{P}$ expressed in terms of
$2k$-fold residue.
The residue formula allows one to derive complicated formulae for the
$\tilde{c}_r(k)$'s ~\cite{CFKRS2}.

Precise information concerning the extreme values
of the zeta function up to given height could be derived from precise knowledge
of the uniform asymptotics of its moments in both the $k$ and $T$
aspects~\cite{FGH}. However, it is not clear how to obtain the uniform
asymptotics from the formulae in~\cite{CFKRS1} and~\cite{CFKRS2}.

As a first step, we decided to try and understand
the behaviour of the coefficients of the moment polynomials.
We hoped that deriving
asymptotics for the coefficients that occur on the random matrix theory side
(i.e. the $c_r(k)$'s) could shed some light on the zeta function side (i.e. the
$\tilde{c}_r(k)$'s). In a sense, that turned out to be the case. In an upcoming
paper~\cite{HR}, we show for a fixed $\alpha <1 $, and uniformly in $0\le
r\le k^{\alpha}$, it holds,
\begin{eqnarray}
    \frac{\tilde{c}_r(k)}{\tilde{c}_0(k)}=(2 {\mathcal B}_k+2\gamma k)^r\,\binom{k^2}{r}\,
    \left[1+O(k^{2(\alpha-1)})\right]\,,\qquad \textrm{as $k\to \infty$}\,,
\end{eqnarray}
where $\gamma=0.5772\ldots$ is Euler's constant, and ${\mathcal B}_k$ is a certain
explicit number (see~\cite{HR} for details) satisfying ${\mathcal B}_k\sim 2k\log k$. To
facilitate comparison, we give the corresponding result for unitary random
matrices (see lemma~\ref{lem:i3}),
\begin{eqnarray}
    \frac{c_r(k)}{c_0(k)}=k^r\,\binom{k^2}{r}\,\left[1+O(k^{2(\alpha-1)})\right]\,,\qquad
    \textrm{as $k\to \infty$}\,.
\end{eqnarray}

\subsection{Basic facts}

For a fixed $r$, one can easily show that $c_r(k)/c_0(k)$ is a polynomial in $k$ of degree $3r$ with leading
coefficient $1/r!$:
\begin{equation}
    \label{eq:c_r as poly}
    b_r(k) := \frac{c_r(k)}{c_0(k)} =  \frac{1}{r!} k^{3r} + \ldots \,.
\end{equation}
Furthermore,  pulling out the $1/r!$, all the coefficients of $r! b_r(k)$ are
polynomials themselves in $\Q[r]$.
These facts can be proven by writing $c_r(k)$, which are
expressible from~\eqref{eq:pkxeq} as elementary symmetric polynomials, in terms of power sums
via Newton's identities
applied to the set of $k^2$ numbers $\{1,2,2,3,3,3,\ldots,2k-2,2k-2,2k-1\}$:
\begin{equation}
    \label{eq:newton}
    r b_r(k) = \sum_{n=1}^r (-1)^{n-1} p_n(k) b_{r-n}(k),
\end{equation}
where $p_n(k)$ are the sums of powers:
\begin{equation}
    \label{eq:power sum}
    p_n(k) = \sum_{j=1}^k j j^n + \sum_{j=k+1}^{2k} (2k-j) j^n.
\end{equation}
The above can be expressed in terms of Bernoulli polynomials using
\begin{equation}
    \label{eq:bernoulli}
    \sum_{j=1}^k j^m = \frac{B_{m+1}(k+1)-B_{m+1}(1)}{m+1}.
\end{equation}
Thus, breaking \eqref{eq:power sum} into three sums, we find
\begin{eqnarray}
    p_n(k) &=& \frac{B_{n+2}(k+1)-B_{n+2}(1)}{n+2}
    + 2k \frac{B_{n+1}(2k)-B_{n+1}(k+1)}{n+1}
    - \frac{B_{n+2}(2k)-B_{n+2}(k+1)}{n+2} \notag \\
    &=&
    \frac{2B_{n+2}(k+1)-B_{n+2}(2k)-B_{n+2}(1)}{n+2}
    + 2k \frac{B_{n+1}(2k)-B_{n+1}(k+1)}{n+1} \notag \\
    &=&
    \frac{2^{n+2}-2}{(n+1)(n+2)} k^{n+2} + \ldots
\end{eqnarray}
is a polynomial in $k$ of degree $n+2$ and leading coefficient $(2^{n+2}-2)/((n+1)(n+2))$.
The last step follows from the expansion of $B_n(x)$ in terms of Bernoulli numbers.

Therefore, from the $n=1$ term in~\eqref{eq:newton}, we inductively get the leading term of~\eqref{eq:c_r as poly},
and the recursion also gives the coefficients of $b_r(k)$ as rational numbers (functions of $r$).
We list the first eight $p_n(k)$:
$p_1(k) = {k}^{3}$,
$p_2(k) = \myfrac{7}{6}\,{k}^{4}-\myfrac{1}{6}\,{k}^{2}$,
$p_3(k) = 3/2\,{k}^{5}-1/2\,{k}^{3}$,
$p_4(k) = {\myfrac {31}{15}}\,{k}^{6}-7/6\,{k}^{4}+1/10\,{k}^{2}$,
$p_5(k) = 3\,{k}^{7}-5/2\,{k}^{5}+1/2\,{k}^{3}$,
$p_6(k) = {\myfrac {127}{28}}\,{k}^{8}-{\myfrac {31}{6}}\,{k}^{6}+7/4\,{k}^{4}-{\myfrac {5}{42}}\,{k}^{2}$,
$p_7(k) = {\myfrac {85}{12}}\,{k}^{9}-21/2\,{k}^{7}+{\myfrac {21}{4}}\,{k}^{5}-5/6\,{k}^{3}$,
$p_8(k) = {\myfrac {511}{45}}\,{k}^{10}-{\myfrac {127}{6}}\,{k}^{8}+
          {\myfrac {217}{15}}\,{k}^{6}-{\myfrac {35}{9}}\,{k }^{4}+{\myfrac {7}{30}}\,{k}^{2}$,
and the first few $b_r(k)$:
        $b_1(k) = {k}^{3}$,
        $b_2(k) = \myfrac{1}{2}\,{k}^{6}-{\myfrac {7}{12}}\,{k}^{4}+\myfrac{1}{12}\,{k}^{2}$,
        $b_3(k) = \myfrac{1}{6}\,{k}^{9}-{\myfrac {7}{12}}\,{k}^{7}+{\myfrac {7}{12}}\,{k}^{5}-\myfrac{1}{6}\,{k}^{3}$,
        $b_4(k) = \myfrac{1}{24}\,{k}^{12}-{\myfrac {7}{24}}\,{k}^{10}+{\myfrac {205}{288}}\,{k}^{8}-{\myfrac {527}{720}}\,{k}^{6}+{
        \myfrac {85}{288}}\,{k}^{4}-\myfrac{1}{40}\,{k}^{2}$,
        $b_5(k) = {\myfrac {1}{120}}\,{k}^{15}-{\myfrac {7}{72}}\,{k}^{13}+{\myfrac {125}{288}}\,{k}^{11}-{\myfrac {677}{720}}
        \,{k}^{9}+{\myfrac {1489}{1440}}\,{k}^{7}-{\myfrac {97}{180}}\,{k}^{5}+\myfrac{1}{10}\,{k}^{3}$,
        $b_6(k) = {\myfrac {1}{720}}\,{k}^{18}-{\myfrac {7}{288}}\,{k}^{16}+{\myfrac {11}{64}}\,{k}^{14}-{\myfrac {32927}{
        51840}}\,{k}^{12}+{\myfrac {22931}{17280}}\,{k}^{10}-{\myfrac {38245}{24192}}\,{k}^{8}+{\myfrac {10513}{
        10368}}\,{k}^{6}-{\myfrac {47}{160}}\,{k}^{4}+{\myfrac {5}{252}}\,{k}^{2}$.

One readily observes from the powers of $k$ that appear that
$b_r(-k)=(-1)^r b_r(k)$,
and this can be proven inductively from the recursion
(first establishing this property for $p_n(k)$ from its expression in terms of
Bernoulli polynomials).

The recursion also allows us to work out specific formulas for the lower terms of
$b_r(k)$. For example, the next to leading term has degree $3r-2$.
Writing
\begin{equation}
    \label{eq:b poly}
    b_r(k) = \sum_{0 \leq j \leq (3r-2)/2} b_{r,j} k^{3r-2j}
\end{equation}
we have, from the $n=1,2$ terms of the recursion, the relation:
    $r b_{r,1} = b_{r-1,1}-7/6 b_{r-2,0}$.
But $b_{r-2,0}=1/(r-2)!$, and one easily checks inductively that
    $b_{r,1} = -\frac{7}{12}\frac{1}{(r-2)!}$.

More generally, writing
    $p_n(k) = \sum_{0 \leq j \leq \frac{n}{2}} p_{n,j} k^{n+2-2j}$,
we have, on comparing the coefficient of $k^{3r-2j}$ on both
sides of~\eqref{eq:newton}, that
\begin{equation}
    \label{eq:coeff recursion}
    r b_{r,j} = \sum_{n=1}^{j+1} (-1)^{n-1}
    \sum_{a=0}^{j+1-n} p_{n,a} b_{r-n,j+1-n-a}.
\end{equation}
One can show, inductively on $j$, that $b_{r,j}$ is of the form
\begin{equation}
     \label{eq:g_j}
     b_{r,j} = \frac{g_j(r)}{r!}
\end{equation}
where $g_j(r)$ is a polynomial in $r$ with rational coefficients,
by plugging this into~\eqref{eq:coeff recursion}, and separating the $(n,a)=(1,0)$
term from the rest as follows. Putting the other terms over a
common denominator, and using $p_{1,0}=1$, we have:
    $r \frac{g_j(r)}{r!} = \frac{g_j(r-1)}{(r-1)!} +\frac{h_j(r)}{(r-1)!}$,
where, $h_j(r)$ is the polynomial obtained by our inductive hypothesis
from the terms $(n,a)\neq(1,0)$. Therefore
    $g_j(r) - g_j(r-1)= h_j(r)$,
and this difference equation allows us to solve for all but the constant
coefficient of $g_j$, which can be obtained from a specific value of $b_{r,j}$.

For example, to work out $b_{r,2}$, we substitute $b_{r,2}=g_2(r)/r!$ into the recursion
    $r b_{r,2} = \sum_{n=1}^{3} (-1)^{n-1} \sum_{a=0}^{3-n} p_{n,a} b_{r-n,3-n-a}$,
and pull out the $(n,a)=(1,0)$ term, giving,
    $r \frac{g_2(r)}{r!} = \frac{g_2(r-1)}{(r-1)!} + \frac{49}{72(r-4)!} + \frac{1}{6(r-2)!} +\frac{3}{2(r-3)!}$.
Clearing denominators we have
    $g_2(r)- g_2(r-1) =  \frac{49}{72}(r-1)(r-2)(r-3) + \frac{1}{6}(r-1) +\frac{3}{2}(r-1)(r-2)
                     = {\myfrac {49}{72}}\,{r}^{3}-{\myfrac {31}{12}}\,{r}^{2}+{\myfrac {227}{72}}\,r-5/4$,
and solving for the coefficients of $g_2$ yields
    $g_2(r) = 49/288 r^4-25/48 r^3 +131/288 r^2 -5/48 r + g_2(0)$.
From $b_{2,2}=g_2(2)/2!=1/12$ we get $g_2(0)=0$. Hence, after factoring the above,
    $b_{r,2} = \frac{49r^2-101r+30}{288(r-2)!}$.
We list the first few terms in the expansion of $b_r(k)$:
{\small
\begin{eqnarray}
    \label{eq:b_r}
    &b_r(k)& = \frac{1}{r!} k^{3r} -\frac{7}{12}\frac{1}{(r-2)!} k^{3r-2} +
             \frac{49r^2-101r+30}{288(r-2)!} k^{3r-4} \notag \\
    &-&\frac {1715\,{r}^{3}-5460\,{r}^{2}+4069\,r-732}{ 51840(r-3)!} k^{3r-6} \notag \\
    &+& \frac {12005\,{r}^{4}-52430\,{r}^{3}+69967\,{r}^{2}-22726\,r-11928}{ 2488320 (r-4)!} k^{3r-8} \notag \\
    &-& \frac{\left( 117649\,{r}^{6}-1130871\,{r}^{5}+3998449\,{r}^{4}-6072801\,{r}^{3}+2402926\,{r}^{2}+383976\, r+6420672 \right)}{209018880(r-4)!} k^{3r-10} \notag \\
    &+& \frac{1}{75246796800(r-5)!}  ( 4117715\,{r}^{7}-49076440\,{r}^{6}+225080030\,{r}^{5}-466558120\,{r}^{4} \notag \\
    && \quad \quad \quad +297681419\,{r}^{3}-108063712\,{r}^{2}+978267588\,r+1471379040 ) k^{3r-12} \notag \\
    &\vdots&
\end{eqnarray}
}

\section{Basic asymptotics}

The coefficients of $P_k(x)$ can be expressed as elementary symmetric
polynomials in its roots. For example, if we let $-z_1,-z_2,\ldots,-z_{k^2}$
denote the roots of $P_k(x)$, then, from the last equation in~\eqref{eq:pkxeq},
we have
    $c_{k^2}(k) = 1$,
    $c_{k^2-1}(k) = \sum_{1\le j\le  k^2} \frac{1}{z_j}$,
    $c_{k^2-2}(k) = \sum_{1\le j_1<j_2\le k^2} \frac{1}{z_{j_1}z_{j_2}}$,
and so on. The question is reduced to estimating
$\sum_{1\le j_1<j_2<\cdots<j_r\le k^2} \frac{1}{z_{j_1}z_{j_2}\ldots z_{j_r}}$.
If $r$ is small, then the terms $z_{j_i}$ in this expression may be thought of
heuristically as independent random variables drawn from the distribution
\begin{eqnarray}
    \mathbb{P}(z=j)=\left\{
    \begin{array}{ll}
        j/k^2\,, & \textrm{if } j\in \{1,\ldots,k\} \\
        \,&\\
        (2k-j)/k^2\,, &\textrm{if } j\in \{k+1,\ldots,2k\}
    \end{array} \right.
\end{eqnarray}
Therefore, for small $r$, one may expect something like the following to hold,
$c_{k^2-r}(k) \approx \binom{k^2}{r}\, \left[\mathbb{E}_P\left(\frac{1}{z}\right)\right]^r = \binom{k^2}{r}\,  \frac{A^r}{k^r}$,
where $A$ is defined in \eqref{eq:adef}.
This is precisely the statement of our lemma~\ref{lem:i1}.

\begin{lem} \label{lem:i1}
Fix $\alpha<1 $. Assume $0\le r \le k^{\alpha}$. Then, uniformly in $r$ over that range,
and as $k\to \infty$, it holds
\begin{eqnarray}
    c_{k^2-r}(k) =\frac{(kA)^{r}}{r!} \left[1+ O(k^{2(\alpha-1)}\log{k})\right]=
    \binom{k^2}{r}\frac{A^r}{k^r}\left[1+O(k^{2(\alpha-1)}\log{k}) \right]\,.
\end{eqnarray}
\end{lem}

\begin{proof}
One may write
\begin{eqnarray} \label{eq:fkxlow}
    P_k(x)&=&\exp\left(\log\, \prod_{j=1}^k (1+x/j)^j \prod_{j=k+1}^{2k} (1+x/j)^{2k-j}\right) \nonumber \\
     &=& \exp\left(\sum_{m=1}^{\infty} \frac{(-1)^{m+1}}{m} \eta_m \,x^m \right)\,,
\end{eqnarray}
where
\begin{eqnarray}
\eta_m:=\sum_{j=1}^k j^{1-m} +\sum_{j=k+1}^{2k}(2k-j)j^{-m},\qquad m\ge 1\,. 
\end{eqnarray}
By the Euler-Maclaurin summation formula, $\eta_1=k\log 4+O(1)$, and $\eta_m \le 2\log k$ for
$m>1$. So, we anticipate the main contribution will come from the term $\exp(\eta_1 x)$. We thus write
\begin{eqnarray}\label{eq:lma1}
\exp(\eta_1 x) \exp\left(\sum_{m=2}^{\infty} \frac{(-1)^{m+1}}{m} \eta_m x^m\right)=: \exp(\eta_1 x)\,f(x)\,,
\end{eqnarray}
and expand $f$ in a Taylor series
\begin{eqnarray} \label{eq:lma2}
    f(x)=1+ \sum_{n=1}^{\infty} \beta_n x^n\,.
\end{eqnarray}
A direct application of Cauchy's estimate yields the upper bound
\begin{eqnarray} \label{eq:low0}
    |\beta_n| = \left|\frac{1}{2\pi i} \int_C \frac{f(z)}{z^{n+1}}\,dz\right| = O(\log(k)^{n/2})\,, 
\end{eqnarray}
where $C$ is the circle centered the origin of radius $1/\sqrt{\log k}$, and
$k$ sufficiently large. Now, $c_{k^2-r}(k)$ is the the coefficient of $x^{r}$
in the Taylor expansion about zero of the function (\ref{eq:lma1}). As
$\beta_1=0$ in expansion (\ref{eq:lma2}), it follows
\begin{eqnarray}
    c_{k^2-r}(k)= \frac{\eta_1^{r}}{r!}+\sum_{n=2}^{r} \frac{\eta_1^{r-n}}{(r-n)!}
    \,\beta_n = \frac{\eta_1^{r}}{r!} \left(1+\sum_{n=2}^{r} \frac{r!}{(r-n)!}\,
    \frac{\beta_n}{\eta_1^n} \right)\,. 
\end{eqnarray}
Fix $\alpha<1$, and let $r\in [0,k^{\alpha}]$. Then estimate (\ref{eq:low0}) yields
$c_{k^2-r}(k)= \frac{\eta_1^{r}}{r!} \left[ 1+O\left(k^{2(\alpha-1)}\log k\right)\right]$,
with the implied constant depending on $\alpha$.
Since $\eta_1=kA$, we arrive at
\begin{eqnarray}
    c_{k^2-r}(k)= \frac{(kA)^{r}}{r!}
    \left[ 1+O(k^{2(\alpha-1)} \log k)\right]\,,
\end{eqnarray}
as required. Finally, to obtain the result in terms of the binomial coefficient, note
\begin{eqnarray}
\frac{A^r}{k^r} \binom{k^2}{r}= \frac{(k A)^r}{r!}\frac{k^2!}{(k^2-r)!\,k^{2r}}=\frac{(kA)^r}{r!}\,\exp\left(\log \prod_{j=0}^{r-1} (1-j/k^2)\right) \,,
\end{eqnarray}
and
\begin{eqnarray}
\exp\left(\log \prod_{j=1}^{r-1} (1-j/k^2)\right)=1+O\left(\sum_{j=1}^{r-1} \sum_{m=1}^{\infty} \frac{j^m}{m\, k^{2m}}\right)=1+O(k^{2(\alpha-1)})\,.
\end{eqnarray}\end{proof}

A similar heuristic applied to the leading coefficients of $P_k(x)$ leads one to expect that
$c_r(k)\approx c_0(k) \binom{k^2}{r} \left[\mathbb{E}_P(z)\right]^r= c_0(k) \binom{k^2}{r}k^r$,
in agreement with lemma~\ref{lem:i3} below.

\begin{lem} \label{lem:i3}
Fix $\alpha <1$. Assume $0 \le r \le   k^{\alpha}$.  Then, uniformly in $r$ over that range, and as $k\to \infty$, it holds
\begin{eqnarray}
\frac{c_{r}(k)}{c_0(k)}=\frac{k^{3r}}{r!} \left[1+O(k^{2(\alpha-1)}) \right]= \binom{k^2}{r} k^r \left[1+O(k^{2(\alpha-1)}) \right]\,.
\end{eqnarray}
\end{lem}

\begin{proof}
The leading coefficients of $P_k(x)$ are the  trailing coefficients of 
$x^{k^2} \,P_k(1/x)=c_0(k) \prod_{j=1}^k (1+jx)^j \prod_{j=k+1}^{2k} (1+jx)^{2k-j}$. 
Also, by Taylor expansions,
$\log \prod_{j=1}^k (1+jx)^j \prod_{j=k+1}^{2k} (1+jx)^{2k-j}= \sum_{m=1}^{\infty} \frac{(-1)^{m+1}}{m} p_m(k) x^m$,
where $p_m(k)=\sum_{j=1}^k j^{m+1}+\sum_{j=k+1}^{2k} (2k-j) j^m$. Now, $c_r(k)/c_0(k)$  is the coefficient of $x^{r}$ in the expansion about zero of
\begin{eqnarray}
    \exp(p_1(k) x) \exp\left(\sum_{m=2}^{\infty} \frac{(-1)^{m+1}}{m} p_m(k) x^m\right)=:\exp(p_1(k) x) \left(1+ \sum_{m=1}^{\infty} \nu_m(k) x^m \right)\,.
\end{eqnarray}
Since $p_m(k)\le (2 k)^{m+2}$ for $m\ge 1$, Cauchy's estimate supplies the
bound $\nu_m(k) =O(k^{2m})$, as $k\to \infty$.  Also, it  easy to verify $p_1(k)=k^3$ and $\nu_1(k)=0$.

Finally, suppose $r\in [0,k^{\alpha}]$, where $\alpha<1$ is a fixed
constant. Then, put together, we have
\begin{eqnarray}
    \label{eq:nu asympt}
    \frac{c_r(k)}{c_0(k)}=\frac{k^{3r}}{r!}+\sum_{m=2}^{r} \frac{k^{3(r-m)}}{(r-m)!} \nu_m(k) =
    \frac{k^{3r}}{r!}\left[1+O(k^{2\alpha-2})\right]\,.
\end{eqnarray}
\end{proof}
By taking more terms in the above lemma, we can obtain a more precise asymptotic,
though still valid only up to $r=O(k^\alpha)$ with $\alpha<1$.
For example, truncating the sum in~\eqref{eq:nu asympt} at $m=5$, gives:
\begin{eqnarray}
    \label{eq:more precise}
    &&\frac{c_r(k)}{c_0(k)}=
    \frac{k^{3r}}{r!}\Biggl(
        1
        -{\frac { \left( r-1 \right) r \left( 7\,{k}^{2}-1 \right) }{12{k}^{4}}} 
        +{\frac { \left( r-2 \right)  \left( r-1 \right) r \left( 3\,{k}^{2}-1 \right) }{6{k}^{6}}} \\
        &+&{\frac { \left( r-3 \right)  \left( r-2 \right)  \left( r-1 \right) r \left( 245 \,{k}^{6}-814\,{k}^{4}+425\,{k}^{2}-36 \right) }{1440{k}^{10}}} \notag \\
        &-&{\frac { \left( r-4 \right)  \left( r-3 \right)  \left( r-2 \right)  \left( r-1 \right) r \left( 105\,{k}^{6}-266\,{k}^{4}+185\,{k}^{2}-36 \right) }{360{k}^{12}}}
        +O\left(k^{6(\alpha-1)}\right)
    \Biggr). \notag
\end{eqnarray}
One can also rearrange this expansion, collecting terms according to the power of
$1/k^2$ to get another derivation of the expansion~\eqref{eq:b_r}.

The situation away from the tails is more complicated. There, we apply a
saddle-point technique in a similar way to~\cite{MW}, where it was used to
develop asymptotics for Stirling numbers of the first kind. Due to such
similarities, some of the details in the proof of lemma~\ref{lem:i2} are only
sketched.

\begin{lem} \label{lem:i2}
Fix $0<\alpha<2$. Let $u:=u_{r,k}$ be the (unique) positive number satisfying
\begin{eqnarray}
    u\sum_{j=1}^k \frac{j}{u+j} + u\sum_{j=k+1}^{2k} \frac{2k-j}{u+j} = k^2-r\,. 
\end{eqnarray}
Further, define $U:=U_{r,k}$ by
\begin{eqnarray}
    U=u \sum_{j=1}^k \frac{j^2}{(u+j)^2}+u\sum_{j=k+1}^{2k}
    \frac{j(2k-j)}{(u+j)^2}\,.
\end{eqnarray}
Then, for $1 < r < k^2- 1$, and as $k\to \infty$, we have
\begin{equation}
    \label{eq:saddle0}
    c_{r}(k) =\frac{P_k(u)}{\sqrt{2\pi U}\, u^{k^2-r}} \left[1+O(1/U) \right]\,.
\end{equation}
If we restrict r to lie in the interval
$k^{\alpha}<r< k^2-k^{\alpha}$, then, as $k\to \infty$,
\begin{equation}
    \label{eq:saddle}
    c_{r}(k) =\frac{P_k(u)}{\sqrt{2\pi U}\, u^{k^2-r}} \left[1+O(k^{-\alpha}) \right]\,.
\end{equation}
\end{lem}

Notice that because we can take $\alpha<2$, the interval covered by this lemma,
$k^\alpha < r < k^2-k^\alpha$, overlaps with the two intervals in
lemmas~\ref{lem:i1} and~\ref{lem:i3}.

\begin{proof}

By Cauchy's theorem,
$c_r(k)=\frac{1}{2\pi i}\int_C \frac{P_k(z)}{z^{k^2-r+1}}\,dz$,
where $C$ is any contour circling the origin once in a positive direction, and
\begin{eqnarray}
P_k(z)=c_0(k)\,\prod_{j=1}^k (z+j)^j\,\prod_{j=k+1}^{2k} (z+j)^{2k-j}\,.
\end{eqnarray}
The plan is to obtain very good approximations of the above integral via a saddle-point method. So, consider the function,
\begin{eqnarray} \label{eq:def11}
f(z):=\log P_k(z)-(k^2-r)\log z\,.
\end{eqnarray}
The saddle points of $f(z)$ are the solutions of
\begin{eqnarray} \label{eq:saddle1}
f'(z)=\sum_{j=1}^k \frac{j}{z+j} +\sum_{j=k+1}^{2k} \frac{2k-j}{z+j}-\frac{k^2-r}{z}=0\,.
\end{eqnarray}
By monotonicity of $z \sum_{j=1}^k j/(j+z)+z\sum_{j=k+1}^{2k} (2k-j)/(j+z)$ for
$z\ge 0$, equation (\ref{eq:saddle1}) has a unique positive solution, which we
denote by $u:=u_{r,k}$. From here on, we follow the standard saddle point
recipe (see~\cite{DB}), in a generally similar way to what was done
in~\cite{MW}. The contour $C$ should cross the saddle-point $u$ in the
direction of steepest descent. The direction of steepest-descent is determined
by the argument of $f''(u)$. Since,
\begin{eqnarray}
    f''(z)=\frac{k^2-r}{z^2}-\sum_{j=1}^k \frac{j}{(z+j)^2} -\sum_{j=k+1}^{2k} \frac{2k-j}{(z+j)^2}\,, 
\end{eqnarray}
then $\textrm{arg } f''(u)=0$ (because $u f''(u)> - f'(u)=0$, as can be seen by showing that
$zf'(z)$ is increasing).
So, as in the standard saddle-point recipe, the angle of passage should be
$\pi/2-(1/2) \textrm{ arg }f''(u)=\pi/2$. This suggests the contour choice
\mbox{$C:=\{ue^{i\theta} : -\pi<\theta<\pi\}$}, which is the same as
in~\cite{MW}. Writing
\begin{eqnarray} \label{eq:int1}
    c_r(k)=\frac{1}{2\pi }\int_{-\pi}^{\pi} e^{f(ue^{i\theta})}\,d\theta\,,
\end{eqnarray}
we have, by calculations analogous to those in~\cite{MW},
\begin{eqnarray} \label{eq:trunc}
c_r(k)=\frac{1}{2\pi}\int_{|\theta|<\epsilon} e^{f(ue^{i\theta})}\,d\theta + O\left(e^{f(u)-c\epsilon^2 U}\right)\,.
\end{eqnarray}
for some absolute constant $c>0$, and $\epsilon>0$ a small number to be chosen later, and
\begin{eqnarray}
    U:=-\left.\frac{d^2}{d\theta^2}f(ue^{i\theta})\right|_{\theta=0}=u \sum_{j=1}^k \frac{j^2}{(u+j)^2}+u\sum_{j=k+1}^{2k} \frac{j(2k-j)}{(u+j)^2}\,.
\end{eqnarray}
Next, expand $f(ue^{i\theta})$ about $\theta=0$ to obtain (for $\theta<1/100$ say)
\begin{eqnarray} \label{eq:exp1}
    f(ue^{i\theta})=f(u)-\frac{U}{2}\theta^2+\sum_{m=3}^{\infty} \gamma_m(u) \theta^m\,. 
\end{eqnarray}
In that expansion, we used the fact $u$ is a saddle point, i.e.
$\left.\frac{d}{d\theta} f(u e^{i\theta})\right|_{\theta=0}=0$. Now, Cauchy's
estimate applied to $\frac{d^2}{dz^2} f(ue^{iz})$ yields $\gamma_m(u) =O(U)$,
for $m\ge 3$. So, choose $\epsilon$ such that,
\begin{eqnarray} \label{eq:secep}
    \epsilon=U^{\delta-1/2},\qquad  0<\delta<1/6 \,,\,\, \textrm{ $\delta$ is fixed}\,.
\end{eqnarray}
 Consequently,
\begin{eqnarray} \label{eq:trunc1}
    c_r(k)=\frac{e^{f(u)}}{2\pi}\int_{|\theta|<\epsilon} e^{-\frac{U}{2}\theta^2}\left(1+\sum_{m=3}^{\infty}\mu_m\theta^m \right)\,d\theta + O\left(e^{f(u)-c U^{2\delta}}\right)\,.
\end{eqnarray}
Cauchy's estimate, in combination with the previous estimate on the $\gamma_m$'s,
gives $\mu_m=O(U^{m/3})$, uniformly in $m$ and $U$. 
(Notice by the assumption $1< r < k^2 -1$ in the statement of the lemma, combined 
 with lemma~\ref{lem:i12}, we have $U > 1.5$ for all $1< r <k^2-1$ and $k$ large enough, and so
  $|\mu_m \theta^m| =O(U^{-(1/6-\delta)m}) = O(1.5^{-(1/6-\delta)m})$. Therefore, 
 the series in \eqref{eq:trunc1} converges geometrically even at extreme values 
  of $r$; i.e. for $r>1$ and 
   $r < k^2-1$.) 
We can also get a sharper
estimate for $m$ fixed.
Exponentiating, $e^{\sum_{m=3}^{\infty} \gamma_m(u) z^m}$,
ensures $\mu_m = O_m(U^{\lfloor m/3 \rfloor})$.

So, if the sum over $m$ in expression (\ref{eq:trunc1}) is
truncated at some integer $M\ge 3$, then, using the first bound on $\mu_m$,
\begin{eqnarray} \label{eq:thr}
c_r(k)&=&\frac{e^{f(u)}}{2\pi}\int_{|\theta|<\epsilon} e^{-\frac{U}{2}\theta^2}\left(1+\sum_{m=3}^{M-1}\mu_m\theta^m\right)\,d\theta \nonumber \\
&&+ O\left( U^{\delta-1/2-(1/6-\delta) M}e^{f(u)}\right)+O\left( e^{f(u)-c U^{2\delta}}\right)\,.
\end{eqnarray}
The domain of integration is extended to $(-\infty,\infty)$, to obtain
\begin{eqnarray} \label{eq:trunc4}
    c_r(k)=\frac{e^{f(u)}}{2\pi}\int_{-\infty}^{\infty} e^{-\frac{U}{2}\theta^2}
    \left(1+\sum_{m=3}^{M-1}\mu_m\theta^m\right)\,d\theta + O\left( U^{\delta-1/2-(1/6-\delta) M}e^{f(u)}\right)\nonumber \\
    +O\left( e^{f(u)-c U^{2\delta}}\right) + O\left(\frac{e^{f(u)}}{2\pi}\int_{|\theta|>\epsilon} 
    e^{-\frac{U}{2}\theta^2}\left(1+\sum_{m=3}^{M-1}\mu_m\theta^m\right)\,d\theta \right)\,.
\end{eqnarray}
The function $\theta^{2m} e^{-U\theta^2/2}$ achieves its maximum at $\theta=\sqrt{2m/U}$.
It follows, on estimating the integrand, that
\begin{eqnarray}
    \int_{\epsilon}^{\infty} \theta^{2m} e^{-\frac{U}{2}\theta^2}\,d\theta =
    O\left(
        \frac{e^{-\frac{U^{2\delta}}{2}}}{U^{(1-2\delta)m}}
    \right)\,, \qquad \textrm{ for $2m<U^{2\delta}$.}
\end{eqnarray}
Thus, if the condition $2M< U^{\delta}$ is imposed in equation (\ref{eq:trunc4}), we obtain
\begin{eqnarray}
    \label{eq:cr saddle}
    c_r(k)&=&\frac{e^{f(u)}}{2\pi}\int_{-\infty}^{\infty} e^{-\frac{U}{2}\theta^2}
    \left(1+\sum_{m=3}^{M-1}\mu_m\theta^m\right)\,d\theta + O\left(U^{\delta-1/2-(1/6-\delta)M}e^{f(u)}\right) \nonumber \\
    &=&\frac{e^{f(u)}}{\sqrt{2\pi U}} \left[1 +\sum_{m=2}^{\lfloor (M-1)/2\rfloor}
    \frac{2^{m}\Gamma(m+1/2)\, \mu_{2m}}{\sqrt{\pi}\, U^{m}} +O\left(U^{\delta-(1/6-\delta) M}\right) \right],
\end{eqnarray}
The first part of the lemma follows by taking $\delta=1/12$, $M=13$ say, and
the estimate $\mu_m = O_m(U^{\lfloor m/3 \rfloor})$.
For the second part we
observe that
\begin{equation} \label{eq:Uxbound}
    U=\Omega(k^{\alpha})  \qquad  \textrm{for}  \qquad   k^{\alpha} < r <
    k^2 - k^{\alpha}
\end{equation}
as $k\to \infty$.
This bound can be proven as follows. First, recall, by definition,
$U(x) = x  \sum_{j=1}^k \frac{ j^2 }{ (j + x)^2 }  +  x
\sum_{j=k+1}^{2k} \frac{ (2k - j) j }{ (j + x)^2 }$.
So, for $x \ge 0$,
    $U(x) \ge x  \sum_{j=1}^k \frac{ j^2 }{ (j + x)^2 }$.
It is not hard to see, for $x \geq k/2$,
    $x \sum_{j=1}^k \frac{ j^2 }{ (j + x)^2 } =
    \Omega(k^3 / x)$.
Also, for $0<x<k/2$,
    $x  \sum_{x < j \le k} \frac{ j^2 }{ (j + x)^2 } = \Omega(xk)$.
By lemma 4, and the monotonicity of the saddle point $u_r$ as a
function of $r$, we have $k^{\alpha -1 } << u_r << k^{3 - \alpha}$ for
$k^{\alpha} < r < k^{2 - \alpha}$.  Combined with the above two lower bounds,
this yields (\ref{eq:Uxbound}).
\end{proof}

\section{A uniform asymptotic}
One unsatisfying feature of lemma~\ref{lem:i2} is the implicit nature of the
asymptotic approximation it provides. Also, it might be desirable to unify
lemmas~\ref{lem:i1} through~\ref{lem:i2} into a single asymptotic approximation
applicable over all regions. The purpose of this section is to resolve these
issues. Lemma~\ref{lem:i12} provides explicit approximations to $u_{r,k}$ and
$U_{r,k}$. Lemma~\ref{lem:i10} provides a uniform asymptotic for $c_r(k)$.

\begin{lem} \label{lem:i12}
Fix $\alpha <2$. Let $u_{r,k}$ and $U_{r,k}$ be as in the statement of
lemma~\ref{lem:i2}, and let $A$ be defined as in (\ref{eq:adef}), so
$A=\log 4+1/(2k)+O(1/k^2)$. Then uniformly in $r$,
\begin{displaymath}
u_{r,k}= \left\{\begin{array}{ll}
\frac{k^2-r}{k A} \left[1+O(k^{\alpha-2}\log k) \right]\,,& \textrm{if $r\in(k^2-k^{\alpha},k^2)$}\,, \\
\,& \\
k^3/r \left[1+O(k^{\alpha-2}) \right]\,,& \textrm{if $r\in(0,k^{\alpha})$}\,.
\end{array}\right. 
\end{displaymath}
$\,$\\
\begin{displaymath}
 U_{r,k}=\left\{\begin{array}{ll}
(k^2-r)\left[1+O(k^{\alpha-2}\log k) \right]\,,& \textrm{if $r\in (k^2-k^{\alpha},k^2)$}\,, \\
\,& \\
r\left[1+O(k^{\alpha-2}) \right]\,,& \textrm{if $r\in(0,k^{\alpha})$}\,. 
\end{array}\right.
\end{displaymath}
\end{lem}

\begin{proof}
Recall the saddle point $u:=u_{r,k}$ is the positive number satisfying
\begin{eqnarray} \label{eq:1001}
u\sum_{j=1}^k \frac{j}{u+j} +u\sum_{j=k+1}^{2k} \frac{2k-j}{u+j}=k^2-r\,. 
\end{eqnarray}
Assume $r \in (k^2 - k^{\alpha} , k^2)$. The left-hand side of equation (\ref{eq:1001}) is strictly increasing for
$u\in (0,\infty)$, so $u_{r,k}$ is unique, and it is easy to see, by direct substitution for $u_{r,k}$, that
\begin{equation}
    \label{eq:u estimate}
    u_{r,k}=O\left((k^2-r)/k\right) = O(k^{\alpha-1}).
\end{equation}
Let us rewrite equation (\ref{eq:1001}) as
$u\sum_{1\le j\le 2u} \frac{j}{u+j}+u\sum_{2u<j\le k} 
\frac{1}{1+u/j}+u\sum_{j=k+1}^{2k} \frac{2k/j-1}{1+u/j}=k^2-r$.
Then,
$u\sum_{1\le j\le 2u} j/(u+j)=O(u^2)$,
$u\sum_{2u<j\le k} 1/(1+u/j)= ku +O(u^2\log k)$, and
$u\sum_{j=k+1}^{2k} (2k/j-1)/(1+u/j)= k u A-ku+O(u^2)$.
Therefore, $uk A+O(u^2\log k)=k^2-r$, and we see, in combination with~\eqref{eq:u estimate}
that
\begin{eqnarray}
u=\frac{k^2-r}{k A} +O\left(\frac{(k^2-r)^2\log k}{k^3} \right)= \frac{k^2-r}{kA}\left[1 +O(k^{\alpha-2}\log k) \right]\,.
\end{eqnarray}
As for the case $r\in (0,k^{\alpha})$, note if $u$ is replaced by $k^3 /(10 r)$, the l.h.s of (3.1) is
\begin{eqnarray}
\frac{k^3}{10r}\sum_{j=1}^k \frac{j}{k^3/(10r)+j} +\frac{k^3}{10r}\sum_{j=k+1}^{2k} \frac{2k-j}{k^3/(10r)+j}\le 
k^2 - 10 r + O(r^2 / k^2) = k^2 - 10 r [ 1 + O(r / k^2) ]\,.
\end{eqnarray}
Therefore, by the monotonicity of the left side of equation (\ref{eq:1001}), the saddle point satisfies $u=\Omega(k^3/r)=\Omega(k^{3-\alpha})$. Using this estimate on $u$,  we can rewrite equation (\ref{eq:1001}) in the form
$\sum_{j=1}^k \frac{j}{1+j/u} +\sum_{j=k+1}^{2k} \frac{2k-j}{1+j/u}=k^2-r$, which implies
$\sum_{j=1}^k j(1-j/u) +\sum_{j=k+1}^{2k} (2k-j)(1-j/u)+O(k^4/u^2)=k^2-r$.
From this, it is straightforward to verify $k^3/u+O\left(k^4/u^2 \right)=r$, or $u=k^3/r\left[1+O(k^{\alpha-2})\right]$, as claimed. Asymptotics for \mbox{$U:=U_{r,k}$} are derived similarly. We have,
\begin{displaymath}
U= \left\{\begin{array}{ll}
uk A+O(u^2\log k)=(k^2-r)\left[1+(k^{\alpha-2} \log k) \right]\,, &\textrm{if $r\in (k^2-k^{\alpha},k^2)$}\,,\\
\\
k^3/u+O\left(k^4/u^2\right) = r\left[1+\left( k^{\alpha-2}\right) \right]\,, &\textrm{if $r\in (0,k^{\alpha})$}\,.
\end{array}\right.
\end{displaymath}
\end{proof}

With the aid of lemma~\ref{lem:i12}, it becomes possible to combine lemmas~\ref{lem:i1} through~\ref{lem:i2} into a single lemma.

\begin{lem} \label{lem:i10}
Let $u:=u_{r,k}$ and $U:=U_{r,k}$ be defined as in the statement of lemma~\ref{lem:i2}. Then, uniformly in $r\in (0,k^2)$, and as $k\to \infty$, it holds
\begin{eqnarray}
    \label{eq:lemi10}
    c_{r}(k) &=&
    \frac{\sqrt{2\pi}\,e^{-k^2}}{k^2!} \binom{k^2}{r}
    \frac{P_k(u)}{\sqrt{U}\, u^{k^2-r}}
    (k^2-r)^{k^2-r+1/2}\,r^{r+1/2}\left[1+O\left(\frac{\log k}{k^{2/3}}\right) \right] \notag \\
    &=&
    \left(\frac{r}{k^2} \right)^r
    \binom{k^2}{r}
    \frac{P_k(u)}{\sqrt{U}\, u^{k^2-r}}
    \left(1-\frac{r}{k^2}\right)^{k^2-r+1/2}\,r^{1/2}\left[1+O\left(\frac{\log k}{k^{2/3}}\right) \right]\,.
\end{eqnarray}
\end{lem}
Note: we exclude $r=0,k$ from the statement of the lemma because the r.h.s.
vanishes at those points.
\begin{proof}
Lemmas \ref{lem:i1}, \ref{lem:i3}, and \ref{lem:i2} give the asymptotics for
$c_r(k)$ in three overlapping regions that cover all $0 \leq r \leq k^2$.
We compare these three asymptotics to the r.h.s. above, and find the $\alpha$
in those lemmas that gives the optimal remainder term (it will transpire that
$\alpha=3/4$).

So fix $\alpha <1 $, and suppose $r\in
(0,k^{\alpha})$. By lemma~\ref{lem:i12}, we have
\begin{eqnarray}
u=k^3/r\,[1+O(k^{\alpha-2})]\,,\qquad U=r[1+O(k^{\alpha-2})]\,.
\end{eqnarray}
\begin{eqnarray}
\Longrightarrow\qquad \log u=\log (k^3/r)+O(k^{\alpha-2})\,,\qquad  \log U=\log r+O(k^{\alpha-2})\,.
\end{eqnarray}
Using $\log(1+x)=x+O(x^2)$ with $x=j/u$, and the above estimate for $u$, we get
\begin{eqnarray}
\log \left(P_k(u)/c_0(k)\right)&=& \sum_{j=1}^k j\log(j+u)+\sum_{j=k+1}^{2k} (2k-j)\log (j+u)\nonumber\\
&=& k^2\log u +\frac{1}{u}\sum_{j=1}^k j^2+\frac{1}{u} \sum_{j=k+1}^{2k} (2k-j)j+O(k^{2\alpha-2})\nonumber\\
&=&k^2\log u +k^3/u+O(k^{2\alpha-2})\,.
\end{eqnarray}
And by Stirling's formula,
\begin{eqnarray}
r!\,\binom{k^2}{r}=\frac{k^2!}{(k^2-r)!}= \frac{k^2!\, e^{k^2-r}}{\sqrt{2\pi}\,(k^2-r)^{k^2-r+1/2}}\left[1+O\left(\frac{1}{k^2-r}\right)\right]\,.
\end{eqnarray}
Put together, for $r\in (0,k^{\alpha})$,
\begin{eqnarray}
&&\frac{\sqrt{2\pi}\,e^{-k^2}}{k^2!} \binom{k^2}{r} \frac{P_k(u)}{\sqrt{U}\, u^{k^2-r}} (k^2-r)^{k^2-r+1/2}\,r^{r+1/2}=\\
&&\frac{c_0(k)}{r!}\,e^{k^2\log u +k^3/u+O(k^{2\alpha-2})-(k^2-r)\log u-(1/2)\log U+(r+1/2)\log r-r}=\notag\\
&&\frac{c_0(k)}{r!}\,e^{r\log (k^3/r)+r-(1/2)\log r+r\log r+(1/2)\log r-r+O(k^{2\alpha-2})} =c_0(k)\,\frac{k^{3r}}{r!}\left[1+O(k^{2\alpha-2})\right] \notag.
\end{eqnarray}
Lastly, by lemma~\ref{lem:i3},
\begin{eqnarray}
\label{eq:range1}
\,\nonumber\\
c_r(k)= \frac{\sqrt{2\pi}\,e^{-k^2}}{k^2!} \binom{k^2}{r} \frac{P_k(u)}{\sqrt{U}\, u^{k^2-r}} (k^2-r)^{k^2-r+1/2}\,r^{r+1/2}\left[1+O\left(k^{2\alpha-2}\right) \right]\,,r\in(0,k^{\alpha}).
\end{eqnarray}
Next, we consider the case $r\in(k^{\alpha},k^2-k^{\alpha})$. By Stirling's formula
\begin{eqnarray}
&&\frac{\sqrt{2\pi}\,e^{-k^2}}{k^2!} \binom{k^2}{r} (k^2-r)^{k^2-r+1/2}\,r^{r+1/2}=\nonumber\\
&&\frac{\sqrt{2\pi}\,e^{-k^2}}{k^2!} \frac{k^2!\,e^{k^2}\,(k^2-r)^{k^2-r+1/2}\,r^{r+1/2}}{2\pi\,(k^2-r)^{k^2-r+1/2}\,r^{r+1/2}} \left[1+O\left(\frac{1}{r}+ \frac{1}{k^2-r}\right)\right]=\nonumber\\
&&\frac{1}{\sqrt{2\pi}}\,\left[1+O\left(k^{-\alpha}\right) \right]\,.
\end{eqnarray}
So by lemma~\ref{lem:i2}, and for $r\in(k^{\alpha},k^2-k^{\alpha})$,
\begin{eqnarray}
\label{eq:range2}
c_r(k)= \frac{\sqrt{2\pi}\,e^{-k^2}}{k^2!} \binom{k^2}{r} \frac{P_k(u)}{\sqrt{U}\, u^{k^2-r}} (k^2-r)^{k^2-r+1/2}\,r^{r+1/2}\left[1+O\left(k^{-\alpha}\right) \right]\,.
\end{eqnarray}
Similar manipulations together with the remark following lemma~\ref{lem:i1} give: for \mbox{$r\in(k^2-k^{\alpha},k^2)$},
\begin{eqnarray}
\label{eq:range3}
c_r(k)= \frac{\sqrt{2\pi}\,e^{-k^2}}{k^2!} \binom{k^2}{r} \frac{P_k(u)}{\sqrt{U}\, u^{k^2-r}} (k^2-r)^{k^2-r+1/2}\,r^{r+1/2}\left[1+O\left(k^{2\alpha-2}\,\log k\right) \right].
\end{eqnarray}
Finally, comparing \eqref{eq:range1}, \eqref{eq:range2}, and \eqref{eq:range3}, we find that
the optimal $\alpha$ satisfies $2-2\alpha=\alpha \Rightarrow \alpha=2/3$, giving
the $O$ term claimed in the lemma. The second formula in the lemma follows from Stirling's formula.

\end{proof}

\section{A more precise asymptotic for the leading coefficients.}
\label{sec:more precise}

In this section we assume that $r=O(k^\alpha)$, with $\alpha<2$, unless otherwise
stated.

As a direct application of the Lagrange inversion formula, one can obtain a more precise expression for the saddle-point $u_{r,k}$ over restricted ranges. This will allow us to write out series
for $u$, $U$, $P_k(u)$, and $\mu_m$ which will give an explicit expansion for the saddle point
approximation.

Recall $u$ is the unique positive number satisfying
\begin{eqnarray} \label{eq:saddle12}
    u\sum_{j=1}^k \frac{j}{u+j} +u\sum_{j=k+1}^{2k} \frac{2k-j}{u+j}=k^2-r\,.
\end{eqnarray}
Since $r<k^\alpha$, we know by lemma~\ref{lem:i12} that $u>2k$ for $k$ large enough. So, we may apply Taylor expansions to the left side of equation (\ref{eq:saddle12}).
We therefore need to solve for $u$ in the equation:
    $\sum_{m=1}^\infty (-1)^{m+1} \frac{p_m(k)}{u^m} = r$
(the $m=0$ term cancels the $k^2$ on the other side of the equation). Substituting $y=k/u$
the last equation becomes
    $\sum_{m=1}^{\infty} (-1)^{m+1} \frac{p_m(k)}{k^{m+2}} y^{m}=\frac{r}{k^2}$.
Now, trivially,
\begin{equation}
    \label{eq:p bound}
    0 < p_m(k) < (2k)^{m+2}\,,
\end{equation}
so that the sum on the l.h.s. above is analytic in
$|y|<1/2$. Furthermore, it has non-vanishing derivative at $y=0$, because $p_1(k)=k^3$.
Using the bound on $p_m(k)$,
for $x>0$ sufficiently small, say $<1/100$, one can find a unique $y>0$, in say $(0,1/10)$,
such that
\begin{equation} \label{eq:series4}
    \sum_{m=1}^{\infty} (-1)^{m+1} \frac{p_m(k)}{k^{m+2}} y^{m}=x\,.
\end{equation}
We can write down the Taylor series for $y$ in terms of $x$ by Lagrange inversion:
\begin{equation} \label{eq:series5}
    y= \sum_{m=1}^{\infty} \lambda_m(k) x ^{m}\,.
\end{equation}
There are two formulas that are useful for finding the $\lambda$'s. We can either substitute
the above formula for $y$ into~\eqref{eq:series4} and compare coefficients of powers of $x$.
Alternatively, we can use the well known formula~(see~\cite{W}, section 5.1):
\begin{equation}
    \lambda_m(k) =\frac{1}{m!}\left.\frac{d^{m-1}}{dy^{m-1}}\, w_k(y)^{-m} \right|_{y=0}\,,
\end{equation}
where $w_k(z)$ is given by
     $w_k(z):=\sum_{m=1}^\infty (-1)^{m+1} \frac{p_m(k)}{k^{m+2}} z^{m-1}$.
Letting
    $a_m = (-1)^{m+1} p_m(k)/k^{m+2}$,
we have
\begin{eqnarray} \label{eq:lambda}
    \lambda_1(k) &=& 1\,, \,
    \lambda_2(k) = -a_2\,, \,
    \lambda_3(k) = 2 a_2^2 -a_3\,, \,
    \lambda_4(k) = 5 a_2 a_3 -a_4 -5 a_2^3 \notag \\
    \lambda_5(k) &=&
    14\,{a_{{2}}}^{4}-21\,{a_{{2}}}^{2}a_{{3}}+6\,a_{{2}}a_{{4}}+3\,{a_{{3}}}^{2}-a_{{5}} \notag \\
    \lambda_6(k) &=&
    -42\,{a_{{2}}}^{5}+84\,{a_{{2}}}^{3}a_{{3}}-28\,{a_{{2}}}^{2}a_{{4}}-28\,a_{{2}}{a_{{3}}}^{2}+7\,a_{
    {2}}a_{{5}}+7\,a_{{3}}a_{{4}}-a_{{6}} \notag \\
    \lambda_7(k) &=&
    132\,{a_{{2}}}^{6}-330\,{a_{{2}}}^{4}a_{{3}}+120\,{a_{{2}}}^{3}a_{{4}}+180\,{a_{{2}}}^{2}{a_{{3}}}^{
    2}-36\,{a_{{2}}}^{2}a_{{5}}-72\,a_{{2}}a_{{3}}a_{{4}} \notag \\
    &&-12\,{a_{{3}}}^{3}+8\,a_{{2}}a_{{6}}+8\,a_{{3}} a_{{5}}+4\,{a_{{4}}}^{2}-a_{{7}}
\end{eqnarray}
Since $y=k/u$, and $x=r/k^2$, we get:
\begin{eqnarray}
    \label{eq:1 over u 2}
    &&\frac{1}{u} = \frac{r}{k^3} \Biggl( 1
    +\,{\frac {7\,{k}^{2}-1}{6{k}^{2}}}\frac{r}{k^2}
    +\,{\frac {22\,{k}^{4}-5\,{k}^{2}+1}{18{k}^{4}}}\left(\frac{r}{k^{2}}\right)^2 \\
    &&+\,{\frac {1357\,{k}^{6}-435\,{k}^{4}+183\,{k}^{2}-25}{1080{k}^{6}}}\left(\frac{r}{k^{2}}\right)^3
    +\,{\frac {4142\,{k}^{8}-1661\,{k}^{6}+1083\,{k}^{4}-359\,{k}^{2}+35}{3240{k}^{8}}}\left(\frac{r}{k^{2}}\right)^4 \notag \\
    &&+\,{\frac {58691\,{k}^{10}-28609\,{k}^{8}+27146\,{k}^{6}-14906\,{k}^{4}+3283\,{k}^{
    2}-245}{45360{k}^{10}}}\left(\frac{r}{k^{2}}\right)^5 \notag \\
    &&+\,{\frac {888146\,{k}^{12}-506685\,{k}^{10}+640353\,{k}^{8}-512890\,{k}^{6}+
    201576\,{k}^{4}-32025\,{k}^{2}+1925}{680400{k}^{12}}}\left(\frac{r}{k^{2}}\right)^6 + \ldots \Biggr)\,. \notag
\end{eqnarray}
Because~\eqref{eq:series5} is analytic in some neighbourhood of $x=0$, we have that there
exists an $\eta>0$ such that the above
converges geometrically for all $r<\eta k^2$,
and, even though the coefficients depend on $k$, does so uniformly in $k$. The latter is
explained by the fact that the l.h.s. of~\eqref{eq:series4} has, by~\eqref{eq:p bound}, non-zero
derivative in some neighbourhood, independent of $k$, of $y=0$, and that one can cover,
independent of $k$, some sufficiently small neighbourhood of $x=0$ by taking $y$ small enough.
Therefore, the inverse function specified in~\eqref{eq:series5} is analytic about $x=0$
in some neighbourhood that is independent of $k$.

We can also write a series for $u$ directly. On reciprocating:
    $u=\frac{k}{\sum_{m=1}^\infty \lambda_m(k) \left(\frac{r}{k^2}\right)^m}$.
Using Maple, we compute:
\begin{eqnarray}
    &&\left(\sum_{m=1}^\infty \lambda_m \,x^{m-1}\right)^{-1} =
    1 - \lambda_{{2}} \,x + \,({\lambda_{{2}}}^{2}-\lambda_{{3}}) \,x^2 \notag \\
    &&- \,(\lambda_{{4}}-2\,\lambda_{{2}}\lambda_{{3}}+{\lambda_{{2}}}^{3}) \,x^3
    + ({\lambda_{{2}}}^{4}-3\,{\lambda_{{2}}}^{2}\lambda_{{3}}+2\,\lambda_{{2}}\lambda_{{4}}+{\lambda_{{3}}}^{2}-\lambda_{{5}}) \,x^4 \notag \\
  &&-\,({\lambda_{{2}}}^{5} -2\,\lambda_{{3}}\lambda_{{4}}+3\,\lambda_{{2}}{\lambda_{{3}}}^{2}
  -4\,{\lambda_{{2}}}^{3}\lambda_{{3}}+3\,{\lambda_{{2}}}^{2}\lambda_{{4}}
  -2\,\lambda_{{2}}\lambda_{{5}}+\lambda_{{6}}) \,x^5 \notag \\
  &&+\,({\lambda_{{2}}}^{6}-5\,{\lambda_{{2}}}^{4}\lambda_{{3}}+4\,{\lambda_{{2}}}^{3}\lambda_{{4}}
  +6\,{\lambda_{{2}}}^{2}{\lambda_{{3}}}^{2}-3\,{\lambda_{{2}}}^{2}\lambda_{{5}}
  -6\,\lambda_{{2}}\lambda_{{3}}\lambda_{{4}}-{\lambda_{{3}}}^{3} \notag \\
  &&+2\,\lambda_{{2}}\lambda_{{6}}
  +2\,\lambda_{{3}}\lambda_{{5}}+{\lambda_{{4}}}^{2}-\lambda_{{7}} ) \,x^6 - \ldots
\end{eqnarray}
Putting the above together gives the first few terms of $u$:
\begin{eqnarray}
    \label{eq:u}
    &&u = \frac{k^3}{r} \Biggl(
        1 -{\frac {7\,{k}^{2}-1}{6 {k}^{2}}} \frac{r}{k^2}
        +{\frac {5\,{k}^{4}-4\,{k}^{2}-1}{36 {k}^{4}}} \left(\frac{r}{k^{2}}\right)^2  \\
        &&+\,{\frac {4\,{k}^{6}+15\,{k}^{4}-24\,{k}^{2}+5}{540 {k}^{6}}} \left(\frac{r}{k^{2}}\right)^3
        +\,{\frac {59\,{k}^{8}-152\,{k}^{6}-114\,{k}^{4}+232\,{k}^{2}-25}{6480 {k}^{8}}}
        \left(\frac{r}{k^{2}}\right)^4 \notag \\
        &&+\,{\frac {92\,{k}^{10}+329\,{k}^{8}-2128\,{k}^{6}+2302\,{k}^{4}-644\,{k}^{2}+49}{27216 {k}^{10}}}
        \left(\frac{r}{k^{2}}\right)^5 \notag \\
        &&+\,{\frac {3101\,{k}^{12}-10620\,{k}^{10}-43827\,{k}^{8}+157640\,{k}^{6}-125649\,{k}^{4}+20580\,{k}^{2}-1225}{1360800{k}^{12}}} \left(\frac{r}{k^{2}}\right)^6\, + \ldots
    \Biggr) \notag
\end{eqnarray}

A similar series for $U$ can be derived from that of $u$. For $u>2k-1$ we have:
\begin{eqnarray}
    U&=&\frac{1}{u} \sum_{j=1}^k \frac{j^2}{(1+j/u)^2}+\frac{1}{u}\sum_{j=k+1}^{2k} \frac{j(2k-j)}{(1+j/u)^2} \notag \\
    &=&
    \frac{1}{u}
        \sum_{j=1}^k j^2 \left(1-\frac{2j}{u}+\frac{3j^2}{u^2}-\ldots \right)
    +\frac{1}{u}
    \sum_{j=k+1}^{2k} j(2k-j) \left(1-\frac{2j}{u}+\frac{3j^2}{u^2}-\ldots \right)
    \notag \\
    &=& \sum_{m=1}^\infty (-1)^{m-1} m p_m(k) u^{-m}\,.
\end{eqnarray}
Substituting $u^{-1}=k^{-1} \sum_{m=1}^\infty \lambda_m(k) (r/k^2)^m$, the above equals
\begin{eqnarray}
    \label{eq:U}
    &&r\sum_{m=1}^\infty (-1)^{m-1} m \frac{p_m(k)}{k^{m+2}} \left(\frac{r}{k^2}\right)^{m-1}
    \left(1+\lambda_2(k) \frac{r}{k^2} +\lambda_3(k) \left(\frac{r}{k^2}\right)^2+\ldots\right) \notag \\
    &=& r \Biggl(
        1-\frac{7k^2-1}{6k^2} \frac{r}{k^2}
        + \,{\frac {5\,{k}^{4}-4\,{k}^{2}-1}{18{k}^{4}}}\left(\frac{r}{k^{2}}\right)^2 \\
        &&-\,{\frac {151\,{k}^{6}-255\,{k}^{4}+129\,{k}^{2}-25}{1080{k}^{6}}} \left(\frac{r}{k^{2}}\right)^3
        + \,{\frac {187\,{k}^{8}-706\,{k}^{6}+168\,{k}^{4}+386\,{k}^{2}-35}{3240{k}^{8}}} \left(\frac{r}{k^{2}}\right)^4 \notag \\
        &&-\,{\frac {1373\,{k}^{10}-9415\,{k}^{8}+18956\,{k}^{6}-14540\,{k}^{4}+3871\,{k}^{2}-245}{45360{k}^{10}}} \left(\frac{r}{k^{2}}\right)^5 \notag \\
        &&+\,{\frac {7777\,{k}^{12}-114000\,{k}^{10}+158361\,{k}^{8}+190960\,{k}^{6}-279813\,{k}^{4}+38640\,{k}^{2}-1925}{680400{k}^{12}}} \left(\frac{r}{k^{2}}\right)^6 - \ldots
    \Biggr) \,. \notag
\end{eqnarray}

Next we work out the series for $\log(P_k(u)/(c_0(k) u^{k^2}))$. Divide the second formula
in~\eqref{eq:pkxeq} by the leading coefficient $c_0(k)$, pull out $u^{k^2}$, expand
each $\log(1+j/u)=j/u-j^2/2u+\ldots$, and sum over $j$ to get:
\begin{eqnarray}
    \label{eq:log P_k}
    \log(P_k(u)/(c_0(k) u^{k^2})) = \sum_{m=1}^\infty \frac{(-1)^{m-1}p_m(k)}{m u^m}\,.
\end{eqnarray}
Substituting
    $\frac{1}{u}=\frac{1}{k}\sum_{m=1}^\infty \lambda_m(k) \left(\frac{r}{k^2}\right)^m$
the above equals
\begin{eqnarray}
    \label{eq:log P_k 2}
     &&r \sum_{m=1}^\infty \frac{(-1)^{m-1}p_m(k)}{m k^{m+2}}\left(\frac{r}{k^2}\right)^{m-1}
     \left(1+\lambda_2(k) \left(\frac{r}{k^2}\right) + \lambda_3(k) \left(\frac{r}{k^2}\right)^2 
     + \ldots \right)^m \notag \\
     &=& r \Biggl( 1 + \frac{7k^2-1}{12k^2}\frac{r}{k^{2}}
         +\frac {13\,{k}^{4}-2\,{k}^{2}+1}{36{k}^{4}} \left(\frac{r}{k^{2}}\right)^2
         +{\frac {583\,{k}^{6}-165\,{k}^{4}+147\,{k}^{2}-25}{2160{k}^{6}}}\left(\frac{r}{k^{2}}\right)^3 \notag \\
         &+&{\frac {1379\,{k}^{8}-428\,{k}^{6}+642\,{k}^{4}-332\,{k}^{2}+35}{6480{k}^{8}}}\left(\frac{r}{k^{2}}\right)^4 \\
         &+&{\frac {3193\,{k}^{10}-1393\,{k}^{8}+3178\,{k}^{6}-2542\,{k}^{4}+637\,{k}^{2}-49}{18144{k}^{10}}}\left(\frac{r}{k^{2}}\right)^5 \notag \\
         &+&{\frac {203849\,{k}^{12}-100470\,{k}^{10}+307587\,{k}^{8}-379420\,{k}^{6}+192639\,{k}^{4}-31710\,{k}^{2}+1925}{1360800{k}^{12}}}\left(\frac{r}{k^{2}}\right)^6
     + \ldots \Biggr). \notag
\end{eqnarray}

Finally, we develop series for $\mu_m$. These are defined by
$1+\sum_{m=3}^\infty \mu_m \theta^m = \exp\left(\sum_{n=3}^\infty \gamma_n(u) \theta^n \right)$,
where $\gamma_n(u)$ are, from~\eqref{eq:exp1}, the Taylor coefficients of $f(ue^{i\theta})$.
Substituting $ue^{i\theta}$ for $z$ in
$f(z)= \log(P_k(z)) - (k^2-r) \log(z)$, we get, after Taylor expanding as
in~\eqref{eq:log P_k} and on expanding $e^{i\theta}$ in its Taylor series, that
\begin{eqnarray}
    \label{eq:gamma_m}
    &&\gamma_n(u) = \frac{(-i)^n}{n!} \sum_{m=1}^\infty
    \frac{(-1)^{m-1}p_m(k)}{u^m}m^{n-1} \\
     &=& \frac{(-i)^n}{n!}
     r \sum_{m=1}^\infty \frac{(-1)^{m-1}p_m(k)m^{n-1}}{k^{m+2}}\left(\frac{r}{k^2}\right)^{m-1}
     \left(1+\lambda_2(k) \left(\frac{r}{k^2}\right) + \lambda_3(k) \left(\frac{r}{k^2}\right)^2
     + \ldots \right)^m\,. \notag
\end{eqnarray}
Now each $\mu_m$ is given as a sum of products of the $\gamma_n(u)$'s, the specific
expression obtained by exponentiating $\sum_3^\infty \gamma_n(u) \theta^n$:
\begin{equation}
    \label{eq:mu in terms gamma}
    \mu_m = \sum_{3j_3 +4j_4 +5j_5 +\ldots =m \atop j_i \geq 0}
    \frac{\gamma_3(u)^{j_3} \gamma_4(u)^{j_4} \gamma_5(u)^{j_5}}{j_3! j_4! j_5! \ldots}\,.
\end{equation}
For example, $\mu_8 =\gamma_8 + \gamma_3\gamma_5 +\gamma_4^2/2$.
Notice that $\gamma_n(u)$ is $r$ times a power series in $r/k^2$ with coefficients that are
polynomials in $1/k^2$.
The latter 
follows from $\lambda_j(k)$ being polynomial a polynomial in $a_m=(-1)^{m+1} p_m(k)/k^{m+2}$,
and $p_m(k)$ being a polynomial in $k$ of degree $m+2$ satisfying,
as described in the introduction, $p_m(k)=(-1)^mp_m(-k)$,
so that the powers of $k$ that appear `go down by twos'.
Therefore, each term in the above sum is of the form $r^a$ times a power series
in $r/k^2$, with coefficients polynomial in $1/k^2$,
and $a \in \Z$, $a\leq \lfloor m/3 \rfloor$.

Substituting~\eqref{eq:u}~\eqref{eq:U}~\eqref{eq:log P_k 2}~\eqref{eq:mu in terms gamma}
and ~\eqref{eq:gamma_m} into the saddle point formula~\eqref{eq:cr saddle},
and taking the logarithm of the various series that appear as factors,
suggests the following formula for $c_r(k)$:
\begin{eqnarray}
    \label{eq:grand explicit}
    \frac{c_r(k)}{c_0(k)} &=& \frac{k^{3r}}{r!} \exp \Biggl(
        -\frac {7r(r-1)}{12k^2}
        -\frac {r(r-1)(26r-15)}{144k^4}
        -\frac {r(r-1)(583r^2-715r+183)}{6480k^6} \notag \\
        &-&\frac {r(r-1)(2758r^3-4499r^2+463r+1491)}{51840k^8} -\ldots 
        - \frac{q_j(r)}{k^{2j}}+\ldots
    \Biggr)\,.
\end{eqnarray}
The $j$-th term appearing in the $\exp$ is of the form $q_j(r)/k^{2j}$
where $q_j(r)$ is a polynomial in $r$ with rational coefficients and
of degree $j+1$.
Furthermore, when $r=0$ or $1$, $c_r(k)/c_0(k) = k^{3r}/r!$, and thus
each term appearing in the $\exp$ must have numerator divisible by $r(r-1)$
so as to vanish at those two values of $r$.

Below we develop our formula further, explaining with proof how it arises.
Substituting the various series as described does not immediately result in
the above formula,
as there are extraneous terms of the form $a_{m,n}/r^mk^{2n}$ along with
various $O$ terms that need to be considered. One also
needs to identify the terms of Stirling's asymptotic formula as appearing
from these substitutions so as to obtain the factor of $r!$
in~\eqref{eq:grand explicit}.

These extraneous terms can be eliminated
by comparing these terms with the terms that appear
in equation~\eqref{eq:b_r}, and considering different values of $r$.

To begin, we have replaced the factors and terms with `just an $r$ and no $k$'
with a $1/r!$. The $P_k(u)/(c_0(k) u^{k^2-r})$ contributes, from~\eqref{eq:log P_k 2}
and~\eqref{eq:u}, a factor of $e^r/r^r$. The $1/\sqrt{U}$ contributes,
from~\eqref{eq:U} a $1/\sqrt{r}$, while the sum involving $\mu_{2m}$
in~\eqref{eq:cr saddle} contributes a factor whose initial terms are
computed to be $\exp(-1/(12r)+1/(360r^3)+1/(1260r^5)-\ldots)$. Together with the
$1/\sqrt{2\pi}$, we recognize these as Stirling's approximation to $1/r!$.
By taking $M$ in~\eqref{eq:cr saddle} sufficiently large, 
letting $k\to \infty$ and $r=k^\epsilon$
with $\epsilon$ sufficiently small, 
and comparing with the asymptotic
$c_r(k)/c_0(k) = k^{3r}/r!(1+O(k^{2(\epsilon-1)}))$ of Theorem~\ref{thm:1},
we have that these terms must coincide with Stirling's
asymptotic series for $1/r!$.

The extraneous terms can be eliminated by comparing with the full polynomial expansion
of $c_r(k)/c_0(k)$.
From equations ~\eqref{eq:b poly} and ~\eqref{eq:g_j},
\begin{equation}
    \label{eq:step 1}
    \frac{r!c_r(k)}{k^{3r}c_0(k)}
    = 1+\sum_{1 \leq j \leq (3r-2)/2} g_j(r) z^j
\end{equation}
where $z=1/k^2$, and $g_j(r)$ is a polynomial in $r$ of degree $\leq 2j$
(the bound on the degree of $g_j(r)$ follows from the inductive procedure for
determining them).
Denote the above polynomial in $z$ by $F_r(z)$. By taking its logarithm,
we can write it in the form
\begin{equation}
    \label{eq:step 2}
    \frac{r!c_r(k)}{k^{3r}c_0(k)} = \exp\left({q}_1(r)z +{q}_2(r)z^2 + {q}_j(r) z^3 + \ldots \right)\,,
\end{equation}
where ${q}_j(r)$ is a polynomial in $r$.
One can easily pass from the polynomial in~\eqref{eq:step 1} to
an expression of the form~\eqref{eq:step 2},
because the polynomial~\eqref{eq:step 1} does not vanish at $z=0$, hence its logarithm
is analytic there and may be expanded in a Taylor series.
The constant term $1$ in~\eqref{eq:step 1} implies that the Taylor series
that appears in the exponent of~\eqref{eq:step 2} has constant term 0,
and thus begins with the ${q}_1(r)z$ term. 
Expanding out $\log{F_r(z)}$ using the Taylor series for
$\log(1+w)=w-w^2/2+w^3/3-\ldots$, and comparing coefficients
with those in the exponent of~\eqref{eq:step 2} shows that
${q}_j(r)$ is a polynomial in $r$ of degree $\leq 2j$.
Our goals are to show that
${q}_j(r)$ actually has degree $j+1$ rather than $\leq 2j$,
and to obtain a truncation bound for~\eqref{eq:step 2}.

To do so, we consider more carefully the result of substituting the various series
into the saddle point formula, pulling out the various factors that comprise
Stirling's asymptotic formula for $r!$, and moving everything else to the $\exp$
by taking the logarithm of the various factors:

First, the series for $\log(P_k(u)/(c_0(k)u^{k^2}))$ is of the form $r$ times a power
series in $r/k^2$, whose $j$th coefficient is polynomial in $1/k^2$ of
degree $j$.

Next, we consider the factor $u^r$. Pulling out, in~\eqref{eq:u}, $k^{3}/r$ 
and moving the rest into the $\exp$, we have that $r\log(u/(k^3/r))$ is of the same form as
$\log(P_k(u)/u^{k^2})$ described in the previous paragraph.

The series for $1/\sqrt{U}$ contributes, on pulling out a $1/\sqrt{r}$, and moving it to the 
$\exp$ as $-\log(U/r)/2$, a power series in $r/k^2$ with coefficients in $k$ as above.

Finally, by the paragraph surrounding~\eqref{eq:mu in terms gamma},
$\mu_{2m}/U^m$ is equal to a sum whose terms are products of the $\gamma_n(u)$'s,
and hence expressible as
power series in $r/k^2$ with coefficients
in $k$ as above, with each term multiplied by a factor of the form $1/r^a$
with $a \geq m-\lfloor 2m/3 \rfloor$. Therefore, 
for $M$ sufficiently large in~\eqref{eq:cr saddle} and also using,
from~\eqref{eq:U}, that $U\sim r$ when $r=o(k^2)$,
we get that the logarithm of 
the bracketed factor in~\eqref{eq:cr saddle} is, 
for any $A>0$, of the form:
\begin{equation}
    \sum_{1\leq a <A} \frac{1}{r^a} \sum_{j=0}^\infty \sigma_{j,a}(k)
    \left(\frac{r}{k^2}\right)^j +O_A(r^{-A})\,,
\end{equation}
where $\sigma_{j,a}(k)$ is a polynomial in $1/k^2$ of degree $j$.

Next, truncate all substituted power series in $x=r/k^2$ after, say, $J=J_{A,\alpha}$ terms so that
the remainder gets absorbed into the $O(r^{-A})$.
This can be achieved because,
by the paragraph following~\eqref{eq:1 over u 2},
all these power series can be shown to be analytic
about $x=0$ with their coefficients, expressed as polynomials in
$1/k^2$, uniformly bounded in $k$.
Hence each has some radius of convergence, and the cost of truncating each power series
at $J$ terms is $O((\eta' r/k^2)^{J+1})$ for some $\eta'>0$ depending on $A$.
Finally, recall that $r=O(k^\alpha)$ with $\alpha<2$. Thus, taking $J$
large enough (depending on $A$ and $\alpha$), we can ensure that the cost
of truncating is $O(r^{-A})$.

Thus, collecting terms according to the power of $1/k^2$, and pulling out
the terms corresponding to Stirling's formula for $r!$ as explained,
we have obtained:
\begin{equation}
    \label{eq:long time coming}
    \frac{r!c_r(k)}{k^{3r}c_0(k)}
    = \exp\left( \sum_{j=1}^J \frac{Q_j(A;r)}{k^{2j}} +O(r^{-A})
    \right)\,,
\end{equation}
where $Q_j(A;r)$ is of the form
\begin{equation}
    \label{eq:Q}
    Q_j(A;r) = \sum_{ -A+j < m \leq j+1} \psi_m(A,j) r^m\,.
\end{equation}
The largest power of $r$ that appears is $j+1$ rather than $j$
because the power series for $\log(P_k(u)/u^{k^2})$ and $r\log(u/(k^3/r))$
are multiplied by an extra $r$. The smallest power is $> -A+j$ rather than $-A$
because $r^m/k^{2j}< r^{m-j}$ can be asborbed into the $O(r^{-A})$ if $m\leq-A+j$.

Formula \eqref{eq:long time coming} essentially gives, as a function of $z=1/k^2$,
the Taylor expansion of $\log(r!c_r(k)/(k^{3r}c_0(k)))$, and comparing
with~\eqref{eq:step 2}, we would like to conclude that $q_j(r)=Q_j(A;r)$,
from which we would have that $q_j(r)$ is a polynomial of degree $j+1$,
rather than of degree $2j$.
However, the presence of extraneous terms and the remainder term
in~\eqref{eq:long time coming} necessitates a more careful comparison,
and we will only conclude that the polynomial part of both
coincide for given $j$ and all $A$ sufficiently large.

From equation~\eqref{eq:nu asympt},
we have that $r!c_r(k)/(k^{3r}c_0(k))$ is asymptotically $1$ when $r=O(k^\alpha)$,
for $\alpha<1$. Therefore its logarithm is analytic in $z=1/k^2$ in that region,
and its Taylor series in~\eqref{eq:step 2} converges in at least that
region.

For $r=O(k^\alpha)$, $\alpha<1$ we can truncate~\eqref{eq:step 2} after sufficiently many
terms so as to also have remainder $O(r^{-A})$. By adjusting our value of $J$ upwards,
we can assume that both series in~\eqref{eq:step 2} and~\eqref{eq:long time coming} 
truncate at $J$. We thus have, on comparing them:
   $\sum_{j=1}^J \frac{q_j(r)-Q_j(A;r)}{k^{2j}}
    = O(r^{-A})$.
We can use this to show that, for any given $j$, and all sufficiently large
$A$, that $q_j(r)=Q_j(A;r)$. We have established that $q_j(r)$ is a polynomial
of degree $\leq 2j$ in $r$, and that $Q_j(A;r)$ has the form~\eqref{eq:Q}. Therefore,
the above can be written as:
   $\sum_{j=1}^J \sum_{-A+j<m\leq 2j} d_{m,j} \frac{r^m}{k^{2j}}
    = O(r^{-A})$.
For each choice of $A$, the l.h.s. above consists of finitely many terms.
If we let $r\sim k^\alpha$, with $\alpha<1$
and $\alpha$ irrational so as to guarantee that all the exponents $m\alpha-2j$ are distinct,
then each term above must individually satisfy the bound:
    $d_{m,j} k^{m\alpha-2j}
    = O(k^{-A\alpha})$.
Therefore, for fixed $j$, $m\geq 0$ and all large enough $A$, we have that $d_{m,j}=0$,
i.e. $q_j(r)$ coincides with the polynomial portion of $Q_j(A;r)$.
This shows that $q_j(r)$ has degree $j+1$.

Next we obtain a bound on the rate of growth of $q_j(r)$. We have seen $q_j(r)$
coincides with the polynomial portion of $Q_j(A;r)$ if we take sufficiently many
terms $m$ in~\eqref{eq:cr saddle}, and have observed that $\mu_{2m}/U^m$ has a factor
of $1/r^a$ with $a=m-\lfloor 2m/3 \rfloor \geq m/3$.
Thus, given $j$, in order for the
various substituted power series to capture the degree $j+1$ polynomial
$q_j(r)$, we need to take, in~\eqref{eq:cr saddle}, $\frac{1}{3}\lfloor (M-1)/2\rfloor \geq j+1$,
so $M\geq6(j+1)$ suffices.
Thus we let $M = 6(j+1)$, and consider:
\begin{equation}
    \label{eq:log saddle point}
    \log \left(
        \frac{r!}{k^{3r}} \frac{P_k(u)}{c_0(k)u^{k^2}}\frac{u^r}{\sqrt{2\pi U}}
        \left[
            1+\sum_{m=2}^{6(j+1)} \frac{2^{m}\Gamma(m+1/2)\, \mu_{2m}}{\sqrt{\pi}\, U^{m}}
        \right]
    \right)\,.
\end{equation}
As before, we substitute the power series in $r/k^2$ for the
various factors and terms of this expression, and letting $z=1/k^2$, express it
as
\begin{equation}
    \label{eq:log 2}
    \sum_{i=1}^\infty Q_{i,j}(r) z^i\,.
\end{equation}
We will obtain a bound for $Q_{i,j}(r)$ and then deduce a bound on $q_j(r)$
by bounding the polynomial part of $Q_{j,j}(r)$. To do so, we regard the above
expansion as a function of a complex variable $r$ and $z$ and apply Cauchy's estimate
twice- first to bound $Q_{i,j}(r)$, and then to bound its polynomial part.

From the power series expansions for
$\log(P_k(u)/(c_0(k)u^{k^2}))$, $r\log(u/(k^3/r))$, and $-\log(U/r)/2$,
we have that:
there exists $a>0$ such that for all $|z|\leq a/|r|$ (think of this as $r/k^2\leq a$),
such that
\begin{equation}
    \label{eq:log 3}
    \log \left(
        \frac{r!}{k^{3r}} \frac{P_k(u)}{c_0(k)u^{k^2}}\frac{u^r}{\sqrt{2\pi U}}
    \right) = O(r).
\end{equation}
Furthermore, in the same kind of region, $|z|< a/|r|$, we have $U=\Omega(|r|)$, and
$\mu_m=O(|r|^{m/3})$, uniformly in $m$.
Combining this with the elementary estimate, $\Gamma(m+1/2)<m^m \sqrt{\pi}$, gives
\begin{equation}
    \label{eq:log 4}
    \log \left(
            1+\sum_{m=2}^{6(j+1)} \frac{2^{m}\Gamma(m+1/2)\, \mu_{2m}}{\sqrt{\pi}\, U^{m}}
        \right) = O(j\log{j}),
\end{equation}
so long as $|r|$ is bounded away from 0.
Therefore, applying Cauchy's estimate to~\eqref{eq:log 2}, taking the circle $|z|=a/|r|$
as the contour of integration, gives
    $Q_{i,j}(r) = O(j\log(j)r(r/a)^i)$.
In particular, $Q_{j,j}(r) = O(j\log(j)r(r/a)^j)$.

To bound the polynomial part $q_j(r)$ of $Q_{j,j}(r)$ we take the integral
\begin{equation}
    \label{eq:poly part}
    q_j(r) = \frac{1}{2 \pi i} \int_C Q_{j,j}(rw)\sum_{m=1}^{j+1} \frac{1}{w^m} dw
\end{equation}
with $C$ the  circle $|w|=2$. Applying the above estimate for $Q_{j,j}(r)$
gives
    $q_j(r) = O(j\log(j)r(r \lambda)^j)$,
for some $\lambda>0$.

Presumably, there should be a more direct and combinatorial derivation
of~\eqref{eq:grand explicit}.
One can verify that expanding~\eqref{eq:grand explicit} using
the Taylor series for $\exp$
does produce the first few terms of~\eqref{eq:b_r}. In fact, by
comparing these two expressions, one can more easily calculate the polynomials
$q_j(r)$ from the terms in~\eqref{eq:b_r}.

\section{A maximal coefficient}

\label{sec:max coeff}

In this section we obtain the estimate for the maximal $c_r(k)$ given in
Theorem~\ref{thm:2}.

The $c_r(k)$'s are the coefficients of a polynomial with negative roots. This implies the sequence $\{c_r(k)\}$ is unimodal (see~\cite{W}). That is, the sequence $\{c_{r}(k)\}$ monotonically increases until it reaches a maximum, then it monotonically decreases. Or there exists $s\in\mathbf{Z}^+$ such that
$c_0(k)\le c_1(k)\le \cdots \le c_s(k)$, and $c_s(k)\ge c_{s+1}(k)\ge \cdots \ge c_{k^2}(k)$.
The idea is to combine unimodality with a saddle-point estimate for $c_{r+1}(k)-c_r(k)$.
Adapting the proof of lemma~\ref{lem:i2}, we consider
\begin{eqnarray} \label{eq:sign1}
    \Delta_r:=c_{r+1}(k)-c_r(k)=\frac{1}{2\pi i} \int_C \frac{(z-1) P_k(z)}{z^{k^2-r+1}}\,dz\,,
\end{eqnarray}
where $C:=\{ue^{i\theta} : -\pi <\theta<\pi\}$. Choose $u:=u_{r,k}$ to be the saddle-point corresponding to $c_r(k)$; that is, $u$ is the positive number satisfying
\begin{eqnarray}
u\sum_{j=1}^k \frac{j}{j+u}+u\sum_{j=k+1}^{2k} \frac{2k-j}{j+u}=k^2-r\,.
\end{eqnarray}

Let $A$ be defined as in (\ref{eq:adef}), so $A=\log 4+1/(2k)+O(1/k^2)$. An
examination of the results in Section 2 reveals a maximal coefficient $c_r(k)$
occurs when the saddle point $u_{r,k}\approx 1$. By lemma~\ref{lem:i12}, this
happens when $k^2-r\approx k A \Longrightarrow r\approx k^2-\mu$. 
Guided by this, our plan is to show, for some $\rho>0$ and
all $k$ sufficiently large, that there exists
integers $ r_1\in (k^2-\mu+\rho \log(k)^2/k,k^2-\mu+1+\rho \log(k)^2/k]$ and
$r_2 \in [k^2-\mu-1-\rho \log(k)^2/k,k^2-\mu-\rho \log(k)^2/k)$,
such that $\Delta_{r_1}$ is strictly negative, and
$\Delta_{r_2}$ is strictly positive. Combined with the unimodality of the
sequence $\{c_r(k)\}$, this shows $c_r(k)$ is maximal for some
$r \in [k^2-\mu-\rho \log(k)^2/k,k^2-\mu+1+\rho \log(k)^2/k]$,
and is not maximal outside of that interval.
To this end, consider the behaviour of $\Delta_r$ for
\begin{eqnarray} \label{eq:asmp1}
    r \in [k^2-kA-2\log(k)^2\,,\,k^2-kA+2\log(k)^2]\,.
\end{eqnarray}
Under assumption (\ref{eq:asmp1}), it follows by lemma~\ref{lem:i12}, 
\begin{eqnarray}\label{eq:simes}
u&:=&u_{r,k}= \frac{k^2-r}{k A}\left[1+O(\log(k)/k) \right]=1+O\left(\frac{\log(k)^2}{k}\right)\,.\nonumber\\
U&:=&U_{r,k}=(k^2-r)\left[1+O(\log(k)/k)\right]=k A+O(\log(k)^2)\,.
\end{eqnarray}
Using the same procedure as in the proof of lemma~\ref{lem:i2}, and choosing $\delta=1/1000$, $M=13$, in expression (\ref{eq:thr}), one obtains
\begin{eqnarray}
    \Delta_r=\frac{e^{f(u)}}{2\pi}\int_{|\theta|<\epsilon} (ue^{i\theta}-1)
    \left(1+\sum_{j=3}^{12} \mu_j \theta^j\right) e^{-\frac{U}{2}\theta^2}\,d\theta
    + O\left(\frac{e^{f(u)}}{U^{5/2}}\right)\,,
\end{eqnarray}
where $\epsilon=U^{1/1000-1/2}$, and $\mu_j = O_j(U^{\lfloor j/3 \rfloor})$. 
Replacing $ue^{i\theta}$ with
$u+iu\theta-u\theta^2/2+O(\theta^3)$ yields
\begin{eqnarray} 
    \Delta_r=\frac{e^{f(u)}}{2\pi}\int_{|\theta|<\epsilon} (u+iu\theta-u\theta^2/2-1) 
    \left(1+\sum_{j=3}^{12} \mu_j \theta^j\right) e^{-\frac{U}{2}\theta^2}\,d\theta + O\left(\frac{e^{f(u)}}{U^{5/2}}\right).
\end{eqnarray}
The integral vanishes for odd powers of $\theta$. Also, the contribution of some the terms can be absorbed into the remainder. For example
\begin{eqnarray}
    \int_{|\theta|<\epsilon} (u-1) \left(1+\sum_{j=3}^{12} \mu_j \theta^j\right) 
    e^{-\frac{U}{2}\theta^2}\,d\theta =\int_{|\theta|<\epsilon} (u-1) e^{-\frac{U}{2}\theta^2}\,d\theta + O\left(U^{-5/2}\right).
\end{eqnarray}
where we made use of $u-1=O((\log^2 k)/k)$, which holds according to (\ref{eq:simes}), 
the bound on $\mu_j$, and also that odd powers of $\theta$ integrate to 0.
After such reductions, the polynomial that survives is $u-1-u\theta^2/2+iu\mu_3\theta^4$. So, when the domain of integration is extended to all of $\mathbb{R}$, we arrive at
\begin{eqnarray} 
\Delta_r=\frac{e^{f(u)}}{2\pi}\int_{-\infty}^{\infty} (u-1-u\theta^2/2+iu\mu_3 \theta^4) e^{-\frac{U}{2}\theta^2}\,d\theta + O\left(\frac{e^{f(u)}}{U^{5/2}}\right)\,.
\end{eqnarray}
Define
\begin{eqnarray}
&&g_m:=\frac{1}{\sqrt{2\pi}}\int_{-\infty}^{\infty} \theta^{2m} e^{-\frac{U}{2} \theta^2}\,d\theta=\frac{U^{-m-1/2}}{\sqrt{2\pi}}\int_{-\infty}^{\infty} \theta^{2m} e^{-\frac{1}{2} \theta^2}\,d\theta\,. \nonumber\\
&&\,\nonumber\\
&&\Longrightarrow \qquad g_0=\frac{1}{U^{1/2}},\qquad g_1=\frac{1}{U^{3/2}},\qquad g_2=\frac{3}{U^{5/2}}\,,
\end{eqnarray}
Then,
\begin{eqnarray} 
\frac{\sqrt{2\pi}\,\Delta_r}{e^{f(u)}}=(u-1)g_0-u g_1/2+i u \mu_3 g_2+ O\left(U^{-5/2}\right)\,.
\end{eqnarray}
Thus, for $k$ greater than some absolute constant and for $r$ in the range (\ref{eq:asmp1}),
\begin{eqnarray} \label{eq:sign201}
    \frac{\sqrt{2\pi U}\,\Delta_r}{e^{f(u)}}= u-1+R\,,
\end{eqnarray}
where
\begin{equation}
    \label{eq:R1}
    R=-\frac{u}{2U} + \frac{3iu\gamma_3}{U^2}+O(U^{-2}).
\end{equation}

Next we obtain a more precise estimate for $\mu_3=\gamma_3$. Differentiating
$f(ue^{i\theta})$ three times with respect to $\theta$ and setting $\theta=0$
gives:
\begin{eqnarray}
    \label{eq:lambda3}
    \gamma_3 &=& \frac{iu}{3!} \sum_{j=1}^k \frac{j^2(u-j)}{(u+j)^3}
    + \sum_{j=k+1}^{2k} \frac{(2k-j)j(u-j)}{(u+j)^3} \notag \\
    &=& -\frac{iu}{3!} (k \log{4} -4u \log{k} + O(1))\,.
\end{eqnarray}
Now, from \eqref{eq:asmp1} and ~\eqref{eq:adef}, we get:
    $\gamma_3 = - \frac{iuU}{3!} +O(\log(k)^2)$.
Therefore, plugging into ~\eqref{eq:R1} and simplifying gives
    $R=\frac{u}{2U}(u-1) + O(U^{-2}(\log k)^2)$
(we have dropped the $O(1/U^2)$ which is subsumed by the $O(U^{-2} (\log k)^2)$).
Hence:
\begin{equation}
    \label{eq:delta}
    \frac{\sqrt{2\pi U}\,\Delta_r}{e^{f(u)}}= (u-1)(1+\frac{u}{2U})+O(U^{-2}(\log k)^2)\,.
\end{equation}
The sign of $\Delta_r$ is, therefore, essentially determined by that of $u-1$.
To obtain the maximal $c_r(k)$ we develop, below, a stronger estimate for $u$
than~\eqref{eq:simes}.

Recall $u$ is the positive number satisfying
\begin{eqnarray}
h(u)=k^2-r, \qquad \textrm{where }\,\,\, h(x):=x\sum_{j=1}^k \frac{j}{j+x}+x\sum_{j=k+1}^{2k} \frac{2k-j}{j+x}\,,
\end{eqnarray}
Also, one may easily verify $U=uh'(u)$. Expanding in a Taylor series,
\begin{eqnarray}\label{eq:tay001}
h(1+x)=h(1)+h'(1) x+O(h''(1)x^2)\,,
\end{eqnarray}
for $|x|<1/2$ say. And by direct calculation,
    $h(1)=\sum_{j=1}^k \frac{j}{j+1}+\sum_{j=k+1}^{2k} \frac{2k-j}{j+1}=k A-\log(k/2)-\gamma+O(1/k)$,
    $h'(1)=\sum_{j=1}^k \frac{j^2}{(1+j)^2} +\sum_{j=k+1}^{2k} \frac{(2k-j)j}{(j+1)^2}
    =k\log{4} +O(\log{k})$, and $h''(1)=O(\log k)$.

Let $\rho>0$, and choose
$d \in (\rho \log(k)^2/k,1+\rho \log(k)^2/k]$ so that $h(1)-d$ is an integer.
Consider the saddle point $u^-$ corresponding to $c_{k^2-h(1)+d}(k)$; i.e.
$h(u^-)=h(1)-d$. By expansion (\ref{eq:tay001}), the point $u^{-}$ satisfies
$u^-=1-d/h'(1)+O\left(\myfrac{(1-u^-)^2\log k}{h'(1)}\right)$, Since
$k^2-h(1)+d$ falls in the range (\ref{eq:asmp1}), it follows from estimate
(\ref{eq:simes}) that $1-u^-=O(\log^5 k/k)$. Consequently,
$u^-=1-d/h'(1)+O(\log(k)^3/k^3)$. Furthermore, by~\eqref{eq:simes}, the
corresponding $U$ satisfies $U\sim kA$. Thus~\eqref{eq:delta} becomes
\begin{equation}
    \label{eq:delta2}
    \frac{\sqrt{2\pi U}\,\Delta_{k^2-h(1)+d}}{e^{f(u)}}= -\frac{d}{h'(1)} + O(k^{-2}(\log k)^2)\,.
\end{equation}
So, choosing $\rho$ sufficiently large, we have, for all sufficiently large $k$, that
$\Delta_{k^2-h(1)+d}$ is strictly negative. Similarly, if one chooses
$d \in [-1-\rho \log(k)^2/k,-\rho \log(k)^2/k)$
we obtain a strict inequality in the opposite direction.
Theorem~\ref{thm:2} follows.

\section{Numerical verifications}
\label{sec:numerics}

Table \ref{table:k 7} compares different approximations to $c_r(7)$.
Recall that $b_r(7):=c_r(7)/c_0(7)$.

Column 3 illustrates that, for smaller $r$, the asymptotics of $b_r(k)$ are simply
described by $k^r\binom{k^2}{r}$, while column 7,
which compares $c_{r}(k)$ to $\binom{k^2}{k^2-r}(A/k)^{k^2-r}$, works
best for $r$ near $k^2$.

Column 4 shows the ratio of $b_r(k)$ to the explicit formula given
in~\eqref{eq:grand explicit}, truncating the terms in the $\exp$
at those that are depicted, namely terms up to $1/k^8$.

Column 5 compares the ratio of $c_r(7)$ to the saddle point formula
\eqref{eq:saddle0}, without the $O$ term. It approximates well throughout, except near
the two tails.

Finally, column 6 shows the performance of the uniform approximation, depicting
$c_r(7)$ divided by the r.h.s. of~\eqref{eq:uniform}, without the $O$-term.
The uniform asymptotic performs surprisingly well across all ranges, and the ratio seems
to equal $1+O(1/k^2)$ uniformly in $r$, rather than
$1+O\left(\log k/ k^{2/3}\right)$ as proven in Lemma~\ref{lem:i10}.

This is somewhat surprising because~\eqref{eq:uniform} was designed to interpolate
three formulas $c_0(k)k^{3r}/r!$, \eqref{eq:saddle0}, and $(kA)^{r}/r!$, across three
asymptotic ranges.
Yet, for every $r$, it seems to approximate better
than any of these. Presumably the uniform approximation correctly
captures lower order terms especially in the tails
where the improvement is substantial.

This is also illustrated by Figure 1 which compares the performance of
the uniform asymptotic to the saddle point asymptotic, for $k=100$.
We see the uniform asymptotic agreeing to within $1/k^2$ across all $r$.

\begin{table}
\begin{footnotesize}
\centerline{
\begin{tabular}{|c|c|c|c|c|c|c|}
$r$ & $b_r(7)$ & $\frac{b_r(7)}{k^r\binom{k^2}{r}}$ & \eqref{eq:grand explicit}  & \eqref{eq:saddle0} &\eqref{eq:uniform} & $\frac{c_{r}(7)}{\binom{k^2}{k^2-r}(A/k)^{k^2-r}}$  \\\hline
1 &    343 &      1 &          1 & 0.922128 & 1.00173 & 0.000359624 \\
2 &  57428 & 0.996599 & 0.9999999993 & 0.959481 & 1.00175 & 0.000472161 \\
3 & 6.25495e+06 & 0.989796 & 0.9999999758 & 0.972666 & 1.00178 & 0.000617783 \\
4 & 4.98350e+08 & 0.979626 & 0.99999979 & 0.97937 & 1.0018 & 0.000805512 \\
5 & 3.09644e+10 & 0.966157 & 0.9999989881 & 0.983419 & 1.00182 & 0.0010466 \\
6 & 1.56208e+12 & 0.94949 & 0.9999964886 & 0.986123 & 1.00184 & 0.00135502 \\
7 & 6.57739e+13 & 0.929761 & 0.9999901402 & 0.988051 & 1.00186 & 0.00174802 \\
8 & 2.35836e+15 & 0.907133 & 0.9999761345 & 0.989491 & 1.00188 & 0.00224681 \\
9 & 7.31054e+16 & 0.8818 & 0.9999482574 & 0.990604 & 1.0019 & 0.00287731 \\
10 & 1.98238e+18 & 0.853982 & 0.9998969593 & 0.991486 & 1.00192 & 0.00367102 \\
11 & 4.74671e+19 & 0.823921 & 0.9998082229 & 0.992198 & 1.00194 & 0.004666 \\
12 & 1.01127e+21 & 0.791877 & 0.9996622029 & 0.992783 & 1.00195 & 0.00590797 \\
13 & 1.92889e+22 & 0.758129 & 0.9994316122 & 0.993268 & 1.00197 & 0.00745151 \\
14 & 3.31096e+23 & 0.722964 & 0.9990798253 & 0.993673 & 1.00199 & 0.00936137 \\
15 & 5.13647e+24 & 0.686679 & 0.9985586683 & 0.994014 &  1.002 & 0.0117138 \\
16 & 7.22761e+25 & 0.649571 & 0.9978058689 & 0.994302 & 1.00202 & 0.0145979 \\
17 & 9.25208e+26 & 0.611939 & 0.9967421379 & 0.994544 & 1.00204 & 0.0181173 \\
18 & 1.08013e+28 & 0.574075 & 0.9952678685 & 0.994747 & 1.00205 & 0.0223911 \\
19 & 1.15237e+29 & 0.536263 & 0.993259448 & 0.994916 & 1.00207 & 0.0275553 \\
20 & 1.12539e+30 & 0.498772 & 0.9905652065 & 0.995055 & 1.00208 & 0.0337638 \\
21 & 1.00737e+31 & 0.461858 & 0.9870010604 & 0.995166 & 1.0021 & 0.0411887 \\
22 & 8.27319e+31 & 0.425756 & 0.9823459696 & 0.995253 & 1.00212 & 0.0500209 \\
23 & 6.23830e+32 & 0.390679 & 0.9763374057 & 0.995315 & 1.00213 & 0.0604688 \\
24 & 4.32068e+33 & 0.356817 & 0.9686671455 & 0.995354 & 1.00215 & 0.0727577 \\
25 & 2.74917e+34 & 0.324337 & 0.9589778584 & 0.995371 & 1.00217 & 0.0871266 \\
26 & 1.60682e+35 & 0.293377 & 0.9468611614 & 0.995366 & 1.00219 & 0.103825 \\
27 & 8.62363e+35 & 0.264051 & 0.9318580758 & 0.995339 & 1.0022 & 0.123107 \\
28 & 4.24711e+36 & 0.236444 & 0.9134631344 & 0.995288 & 1.00222 & 0.145227 \\
29 & 1.91769e+37 & 0.210618 & 0.8911337471 & 0.995212 & 1.00224 & 0.170425 \\
30 & 7.92900e+37 & 0.186607 & 0.8643068022 & 0.995109 & 1.00226 & 0.198924 \\
31 & 2.99741e+38 & 0.164424 & 0.8324247939 & 0.994977 & 1.00228 & 0.230911 \\
32 & 1.03405e+39 & 0.144058 & 0.794973906 & 0.994812 & 1.00231 & 0.266526 \\
33 & 3.24796e+39 & 0.12548 & 0.751536253 & 0.994609 & 1.00233 & 0.305843 \\
34 & 9.26348e+39 & 0.108643 & 0.701857615 & 0.994363 & 1.00236 & 0.348854 \\
35 & 2.39123e+40 & 0.0934815 & 0.6459301241 & 0.994064 & 1.00238 & 0.395449 \\
36 & 5.56518e+40 & 0.0799208 & 0.5840860639 & 0.993702 & 1.00241 & 0.445396 \\
37 & 1.16242e+41 & 0.0678739 & 0.5170938654 & 0.993262 & 1.00244 & 0.498322 \\
38 & 2.16719e+41 & 0.0572458 & 0.4462404757 & 0.992724 & 1.00247 & 0.553696 \\
39 & 3.58295e+41 & 0.0479358 & 0.3733761815 & 0.992059 & 1.0025 & 0.610815 \\
40 & 5.21118e+41 & 0.0398399 & 0.3008906013 & 0.991225 & 1.00253 & 0.668788 \\
41 & 6.60301e+41 & 0.0328524 & 0.2315856915 & 0.99016 & 1.00256 & 0.726538 \\
42 & 7.20033e+41 & 0.0268682 & 0.1684190354 & 0.988765 & 1.00259 & 0.782801 \\
43 & 6.65250e+41 & 0.0217843 & 0.1141152036 & 0.986875 & 1.00261 & 0.836136 \\
44 & 5.10155e+41 & 0.0175011 & 0.07068892633 & 0.984198 & 1.00261 & 0.88495 \\
45 & 3.15679e+41 & 0.0139236 & 0.03898646968 & 0.980152 & 1.0026 & 0.92753 \\
46 & 1.51291e+41 & 0.0109627 & 0.01840875972 & 0.97341 & 1.00254 & 0.962089 \\
47 & 5.26306e+40 & 0.00853539 & 0.00698803733 & 0.960117 & 1.00242 & 0.986828 \\
48 & 1.18076e+40 & 0.0065654 & 0.001898534367 & 0.922532 & 1.00217 &      1 \\
49 & 1.28039e+39 & 0.00498356 & 0.0002772583691 & 2.05827e-05 &    Inf &      1 \\
\hline
\end{tabular}
}
\caption{A comparison of five formulas for the asymptotics of $c_r(7)$, as explained in
Section~\ref{sec:numerics}.
}\label{table:k 7}
\end{footnotesize}
\end{table}

\begin{table}
\begin{footnotesize}
\centerline{
\begin{tabular}{|c|c|c|c||c|c|c|c|}
$k$ & $r$ & $k^2-\mu$ &$r-k^2+\mu$ &
$k$ & $r$ & $k^2-\mu$ &$r-k^2+\mu$ \\ \hline
2 & 3 & 2.5833333 & 0.41666667 &	51 & 2535 & 2534.6271 & 0.37291935\\
3 & 7 & 6.5166667 & 0.48333333 &	52 & 2637 &  2636.26 & 0.74002916\\
4 & 13 & 12.372619 & 0.62738095 &	53 & 2740 & 2739.8925 & 0.1075001\\
5 & 20 & 20.181349 & -0.18134921 &	54 & 2846 & 2845.5247 & 0.47531881\\
6 & 30 & 29.958261 & 0.041738817 &	55 & 2954 & 2953.1565 & 0.84347267\\
7 & 42 & 41.712279 & 0.28772061 &	56 & 3063 & 3062.7881 & 0.21194975\\
8 & 56 & 55.449036 & 0.55096431 &	57 & 3175 & 3174.4193 & 0.58073872\\
9 & 72 & 71.172312 & 0.82768841 &	58 & 3289 & 3288.0502 & 0.94982886\\
10 & 89 & 88.884769 & 0.11523121 &	59 & 3404 & 3403.6808 & 0.31920997\\
11 & 109 & 108.58835 & 0.41164848 &		60 & 3522 & 3521.3111 & 0.68887239\\
12 & 131 & 130.28452 & 0.71547681 &		61 & 3641 & 3640.9412 & 0.0588069\\
13 & 154 & 153.97441 & 0.025587194 &	62 & 3763 & 3762.571 & 0.42900475\\
14 & 180 & 179.65891 & 0.34109033 &		63 & 3887 & 3886.2005 & 0.79945758\\
15 & 208 & 207.33873 & 0.66127328 &		64 & 4012 & 4011.8298 & 0.17015746\\
16 & 237 & 237.01444 & -0.01444432 &	65 & 4140 & 4139.4589 & 0.54109679\\
17 & 269 & 268.68654 & 0.31345885 &		66 & 4270 & 4269.0877 & 0.91226833\\
18 & 303 & 302.35542 & 0.64458331 &		67 & 4401 & 4400.7163 & 0.28366517\\
19 & 338 & 338.02141 & -0.021407767 &	68 & 4535 & 4534.3447 & 0.65528068\\
20 & 376 & 375.6848 & 0.315199 &	69 & 4670 & 4669.9729 & 0.02710854\\
21 & 416 & 415.34584 & 0.65415768 &		70 & 4808 & 4807.6009 & 0.3991427\\
22 & 457 & 457.00474 & -0.0047443102 &	71 & 4948 & 4947.2286 & 0.77137735\\
23 & 501 & 500.66169 & 0.33830803 &		72 & 5089 & 5088.8562 & 0.14380693\\
24 & 547 & 546.31685 & 0.68315269 &		73 & 5233 & 5232.4836 & 0.51642612\\
25 & 594 & 593.97035 & 0.029647012 &	74 & 5379 & 5378.1108 & 0.88922979\\
26 & 644 & 643.62234 & 0.37766471 &		75 & 5526 & 5525.7378 & 0.26221304\\
27 & 696 & 695.27291 & 0.72709347 &		76 & 5676 & 5675.3646 & 0.63537115\\
28 & 749 & 748.92217 & 0.077832963 &	77 & 5827 & 5826.9913 & 0.0086995789\\
29 & 805 & 804.57021 & 0.42979318 &		78 & 5981 & 5980.6178 & 0.38219397\\
30 & 863 & 862.21711 & 0.78289307 &		79 & 6137 & 6136.2441 & 0.75585013\\
31 & 922 & 921.86294 & 0.13705943 &		80 & 6294 & 6293.8703 & 0.12966403\\
32 & 984 & 983.50777 & 0.49222584 &		81 & 6454 & 6453.4964 & 0.50363176\\
33 & 1048 & 1047.1517 & 0.84833194 &	82 & 6616 & 6615.1223 & 0.87774959\\
34 & 1113 & 1112.7947 & 0.20532266 &	83 & 6779 & 6778.748 & 0.2520139\\
35 & 1181 & 1180.4369 & 0.56314762 &	84 & 6945 & 6944.3736 & 0.62642122\\
36 & 1251 & 1250.0782 & 0.92176065 &	85 & 7112 & 7111.999 & 0.00096817198\\
37 & 1322 & 1321.7189 & 0.28111928 &	86 & 7282 & 7281.6243 & 0.37565153\\
38 & 1396 & 1395.3588 & 0.64118441 &	87 & 7454 & 7453.2495 & 0.75046815\\
39 & 1471 & 1470.9981 & 0.0019199218 &	88 & 7627 & 7626.8746 & 0.12541502\\
40 & 1549 & 1548.6367 & 0.3632924 &		89 & 7803 & 7802.4995 & 0.5004892\\
41 & 1629 & 1628.2747 & 0.72527087 &	90 & 7981 & 7980.1243 & 0.87568788\\
42 & 1710 & 1709.9122 & 0.087826568 &	91 & 8160 & 8159.749 & 0.25100833\\
43 & 1794 & 1793.5491 & 0.45093271 &	92 & 8342 & 8341.3736 & 0.62644789\\
44 & 1880 & 1879.1854 & 0.81456437 &	93 & 8525 & 8524.998 & 0.0020040182\\
45 & 1967 & 1966.8213 & 0.17869823 &	94 & 8711 & 8710.6223 & 0.37767423\\
46 & 2057 & 2056.4567 & 0.54331253 &	95 & 8899 & 8898.2465 & 0.75345612\\
47 & 2149 & 2148.0916 & 0.90838688 &	96 & 9088 & 9087.8707 & 0.12934738\\
48 & 2242 & 2241.7261 & 0.27390216 &	97 & 9280 & 9279.4947 & 0.50534575\\
49 & 2338 & 2337.3602 & 0.63984041 &	98 & 9474 & 9473.1186 & 0.88144904\\
50 & 2435 & 2434.9938 & 0.006184772 &	99 & 9669 & 9668.7423 & 0.25765515\\
\hline
\end{tabular}
}
\caption{A comparison of the maximal $c_r(k)$ for $2 \leq k \leq 99$,
to the value $k^2-\mu$ of Theorem~\ref{thm:2}. The Theorem states that, there exists
$\rho>0$ such that, for all $k$ sufficiently
large, their difference lies in $[-\rho \log{k}^2/k, 1+\rho \log{k}^2/k]$.
}\label{table:max coeff}
\end{footnotesize}
\end{table}

\begin{figure}
    \centerline{
        \includegraphics[width=5.2in]{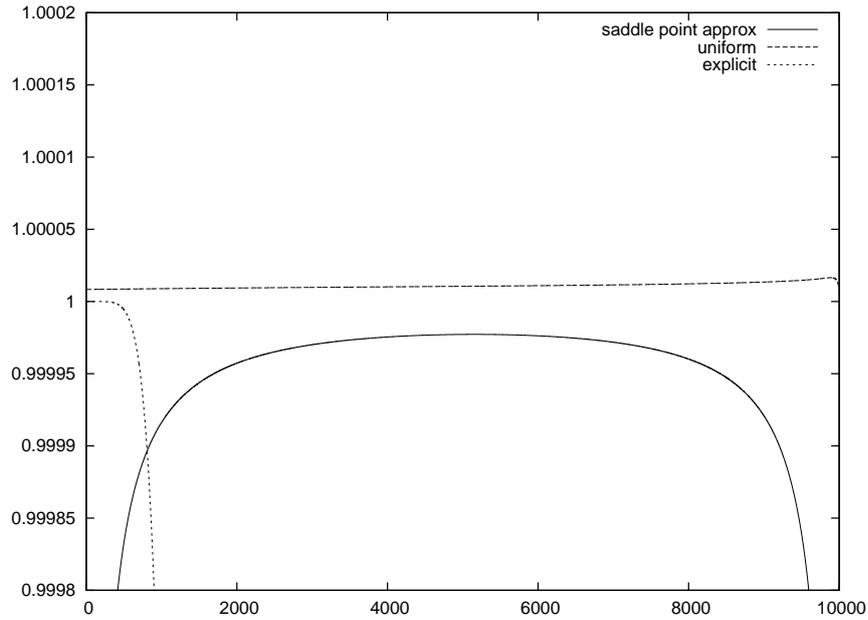}
    }
    \caption{We compare three approximations for $c_r(100)$, $0 < r < 10000$. The long-dashed line
    depicts the ratio of $c_r(100)$ to the r.h.s. of the uniform
    approximation~\eqref{eq:uniform}, while
    the solid line shows the ratio of $c_r(100)$ to the r.h.s. of the saddle point approximation~\eqref{eq:saddle0},
    in both cases without the $O$-term. The short-dashed line plots the ratio of $c_r(100)/c_0(100)$ to the
    formula shown in~\eqref{eq:grand explicit}, taking the first 4 terms in the
    $\exp$, i.e. dropping terms with a $1/k^{10}$ or higher.
    }
    \label{fig:k 100}
\end{figure}

\end{document}